\documentclass[12pt,preprint]{aastex}

\newcommand\iso[2]{$^{\rm #1}$#2}

\def\ciso{\mbox{$^{12}$C/$^{13}$C}}
\def\mv{\mbox{$M_{\rm V}$}}
\def\vk{\mbox{$V-K$}}
\def\teff{\mbox{$T_{\rm eff}$}}
\def\logg{\mbox{log~{\it g}}}
\def\vmicro{\mbox{$\xi_{\rm t}$}}
\def\kmsec{\mbox{km~s$^{\rm -1}$}}

\slugcomment{to be published in AJ}

\shorttitle{CHEMICAL COMPOSITIONS OF FIELD RHB STARS}
\shortauthors{Af{,\s}ar et al.}

\begin{document}

\title{CHEMICAL COMPOSITIONS OF THIN-DISK, HIGH-METALLICITY 
RED HORIZONTAL-BRANCH FIELD STARS}

\author{M. Af\c{s}ar\altaffilmark{1,2}, C. Sneden\altaffilmark{2},
        and B.-Q. For\altaffilmark{2,3}}

\altaffiltext{1}{Department of Astronomy and Space Sciences, 
                 Ege University, 35100 Bornova, \.{I}zmir, Turkey; 
                 melike.afsar@ege.edu.tr}

\altaffiltext{2}{Department of Astronomy and McDonald Observatory,
                 The University of Texas, Austin, TX 78712;
                 afsar,chris,biqing@astro.as.utexas.edu}

\altaffiltext{3}{ICRAR, University of Western Australia, 35 Stirling Hwy,
                 Crawley, WA 6009, Australia; biqing.for@uwa.edu.au}


\begin{abstract}

We present a detailed abundance analysis and atmospheric parameters 
of 76 stars from a survey to identify field Galactic red 
horizontal-branch (RHB) stars. 
High-resolution echelle spectra (R~$\simeq$~60,000, S/N~$\geq$~100) were 
obtained with 2.7~m Smith Telescope at McDonald Observatory. 
The target stars were selected only by color and parallax information.
Overall metallicities and relative abundances of proton-capture elements
(\ion{C}{1}, \ion{N}{1}, \ion{O}{1}, \ion{Li}{1}), 
$\alpha$-elements (\ion{Ca}{1} and \ion{Si}{1}), and neutron-capture 
elements (\ion{Eu}{2} and \ion{La}{2}) were determined by either 
equivalent width or synthetic spectrum analyses.
We used CN features at $\lambda\lambda$7995$-$8040 region in order to 
determine \ciso~ratios of our targets. 
Investigation of the evolutionary stages, using spectroscopic \teff\ 
and \logg~values along with derived \ciso~ratios, revealed the 
presence of 18 probable RHB stars in our sample. 
We also derived kinematics of the stars with available distance information. 
Taking into account both the kinematics and probable evolutionary stages, 
we conclude that our sample contains five thick disk and 13 thin 
disk RHB stars. 
Up until now, RHB stars have been considered as members of the thick 
disk, and were expected to have large space velocities and sub-solar 
metallicities. 
However, our sample is dominated by low velocity solar-metallicity RHB 
stars; their existence cannot be easily explained with standard 
stellar evolution.

\end{abstract}


\section{INTRODUCTION\label{intro}}

In conjunction with onset of quiescent helium fusion, stars locate 
on the zero-age horizontal branch (HB) of the HR diagram.
Usually HB stars have two energy sources:  in addition to the helium burning 
in their cores, they experience hydrogen fusion in a surrounding shell. 
The HB domain encompasses a very large effective temperature range,
leading to stars being labeled as members of the extreme HB, blue HB, 
RR Lyrae variables, 
red HB (RHB), and red clump (RC) stars.
The loci of the stars on HB partly depends on many parameters, including
stellar mass, metallicity, age, helium abundance, and rotation, and mass.
Mass is a key parameter: HB locations depend sensitively on initial 
stellar masses or the amount of mass loss encountered during 
the red giant branch (RGB) phase. 
Theoretical models (e.g. \citealt{swe78}, \citealt{pie04}) 
suggest that very low mass stars (around 0.5~M$_\sun$) appear on the blue 
part of the HB, higher-mass ones (around 0.8 M$_\sun$) sit on the 
RHB, and stars of solar mass and greater are located in the small RC area
at the extreme red edge of the HB.

RHBs are located between the instability strip and the RC.
Stars of the RC
concentrate at \mv~$\thickapprox$~1 and \bv~$\thickapprox$~0.9-1 
(\citealt{can70}, \citealt{wil76}).
These colors correspond to effective temperatures 
\teff~$\thickapprox$~5000--4800~K (e.g, \citealt{ram05}), in contrast 
to the color/\teff\ range of RHB stars ($\bv$)$_{0}~\thickapprox$~0.5--0.8,
\teff~$\thickapprox$~5000--6200~K (\citealt{str81}, \citealt{gra09}).
It is easy to identify RHB stars in globular cluster (GC)
color-magnitude diagrams (e.g, see the survey of \citealt{ros00a}, 
\citealt{ros00b}) but they are not easily distinguished from RC, SGB 
(subgiant branch), and even MS (main sequence) stars in 
the Galactic field. 
There have been several attempts to isolate field RHB stars. 
An early systematic survey was that of \cite{str81}, who observed
eight RHB candidates with the Vilnius photometric system. 
They derived the metallicities and determined the luminosity classes 
using color-color diagrams, but could not prove their RHB membership 
because of lack of surface gravity information. 

Field RHB stars were first investigated as a group by \cite{ros85}.
He studied G5-G7 stars selected from the \cite{upg62,upg63} North 
Galactic Pole (NGP) survey. 
By analyzing low resolution spectroscopic data, Rose showed that one
can distinguish between evolved and ``post-main sequence stars'' (i.e,
subgiants) using the strengths of \ion{Sr}{2} $\lambda$4077 {\AA} and 
CN $\lambda$3883 and 4216 {\AA} bands. 
He also estimated the scale height of RHBs to be $>$ 500~pc and concluded 
that they are moderately metal-poor members of the ``thick disk" 
(\citealt{gil83}) of the Galaxy.

The RHB assignments subsequently were questioned by \cite{nor87}, 
who reported the results of photometric observations of ten 
\cite{upg62,upg63} stars that formed part of the \cite{ros85}
RHB candidate sample.
Norris argued that the colors of these stars 
(e.g., \bv~$\thickapprox$~0.88)
are indistinguishable from the RC stars in the old disk, mildly 
metal-deficient open cluster NGC 2243 ($t$~$\thickapprox$~4~Gyr,
\citealt{ant05}; [Fe/H]~$\thickapprox$~$-$0.4, \citealt{jac11}). 
In fact, the colors of the Rose candidates have almost no 
overlap with the colors of true RHB stars 
(0.6~$\lesssim$~\bv~$\lesssim$~0.8; e.g., \citealt{hes87})
of the globular cluster 47~Tuc ($t$~$\thickapprox$~13~Gyr,
\cite{ant05}; [Fe/H]~=~$\thickapprox$~$-$0.7, \cite{hes87}).

The evolutionary states of alleged RHB field stars were 
revisited by \cite{tau96}, who carried out Vilnius system photometric 
observations of 13 field RHB candidates from \cite{ros85}.
From these data she deduced spectral types, atmospheric 
parameters and absolute magnitudes.
Her results yielded a complex message:  low mean metallicity 
($<$[Fe/H]$>$~$\thickapprox$ $-$0.6) and large mean age 
($t$~$\thickapprox$~10-12 Gyr) consistent with possible RHB status, 
but low mean temperature ($<$\teff$>$~$\thickapprox$~4940~K) consistent 
with the RC.

In another study, \cite{tau97} gathered high-dispersion spectra of 10 
field RHB stars that had been identified in a number of previous papers. 
The metallicity range of her sample was very large, 
$-$0.2~$\geq$~[Fe/H]~$\geq$~$-$1.9, which resulted in mixed Galactic
population membership.
She concluded that the chemical compositions of these RHB stars
were in accord with those of dwarf and red giant stars of similar 
metallicities, and thus RHB stars can be used as tracers of the 
chemical evolution of the Galaxy.
\cite{tau01} followed this paper with another chemical composition
study of \cite{ros85} stars, finding in particular:
\textit{(a)} overabundances of $\alpha$-elements (O, Mg, Si, Ca, Ti);
\textit{(b)} solar abundances of some neutron-capture elements (Y, Ba, La)
and overabundances of others (especially Eu, possibly Zr and Sm):
\textit{(c)} depleted C and enriched N abundances, indicative of
CN-cycle H-fusion in these stars;
and \textit{(d)} perhaps most importantly, very low carbon isotope
ratios, \ciso~$\thickapprox~$3-6.

Previous studies have suggested that true field RHB stars
are relatively rare, but the extant sample sizes are small.
The ones that exist should be relatively low mass members of the 
thick disk because stellar evolution models (e.g. \citealt{lee94}) 
predict that higher-mass, solar-metallicity core He-burning stars 
should reside exclusively in the RC.
Following the \cite{nor87} suggestion that, ``the only way to resolve
this problem in an unambiguous way will be to analyze a complete
sample of stars having colors in the range 0.7~$\leq$~\bv~$\leq$~0.95'',
\cite{kae05} identified a large sample of RHB candidates from the 
Hipparcos catalog (\citealt{per97}).
Their kinematic analysis found both thick disk and halo RHBs, and
they estimated a Galactic scale height of 0.6 kpc for the thick disk RHBs.

In this paper we report on a new large-sample chemical composition 
study of candidate field RHB stars. 
The combined kinematics, metallicities, and chemical abundance
ratios will be used to confirm the existence of a substantial number of 
high-metallicity thin-disk RHB stars.
In \S\ref{obs} we present the target star selection and the spectroscopic
data that were gathered.
Model atmosphere parameter derivation is described in \S\ref{params}.
We discuss the chemical composition analysis in \S\ref{abunds}, 
along with comparison to other large-sample literature studies.
The kinematics of our sample are given in \S\ref{motions}, and
estimates of their evolutionary states in \S\ref{evstat}.
We discuss the implications of these results in \S\ref{interp}.

\section{OBSERVATIONS\label{obs}}

We obtained high resolution, high signal-to-noise spectra of
stars selected mainly by their colors (as suggested above) and
some indications that they might be giants (spectral luminosity 
class III and/or low \mv\ from Hipparcos parallax).

\subsection{Selection of the Program Stars\label{select}}

We assembled our program star list from several sources.
Most of our RHB candidates were selected from among the \cite{upg62}
NGP survey and the \cite{har69,har74,har81} stars used by \cite{ros85}.  
We also made use of 
SIMBAD\footnote{http://simbad.u-strasbg.fr/simbad/} and 
Vizier\footnote{http://vizier.u-strasbg.fr/viz-bin/VizieR}.

We restricted the prospective sample in color, spectral type,
and apparent magnitude (V $<$ 11). 
For spectral type, we limited our search to between G0III and G8III. 
The reason for this restriction is simple: the loci of the RHB stars 
in the HR diagram fall in between the instability strip on the blue side
and RC region on the red side. 
RHB stars have very similar luminosities to RC stars, 
although the effective temperatures of RCs are lower.  
For RC stars \cite{kae05} adopted a luminosity-dependent color 
range of 0.85~$\lesssim$~\bv~$\lesssim$~1.2 (spectral classes
approximately G6--K3), and for RHB stars 
0.5~$\lesssim$~\bv~$\lesssim$~0.8 (approximately G0--G5);
see their Figure~1.
In our color selection we used this RHB \bv\ range, but when
possible we preferred to use an equivalent \vk\ range of 
$\thickapprox$1.5--2.2, because \vk\ colors are almost independent of 
metallicity and gravity but very sensitive to effective temperature.
In order to calculate photometric temperatures
we have used the metallicity-dependent \teff-color formula 
for giants given by \cite{ram05}. 
The apparent magnitude limit was defined by our telescope/instrument
combination; see \S\ref{data}.
The radial velocity and proper motion properties of the 
candidates were not included in the selection criteria, which allowed us 
to minimize potential kinematic biases in our sample. There 
is an inevitable bias due to the cutoff in apparent 
magnitude limit of the instrumental set-up.

\subsection{Observations and data reduction\label{data}}

We obtained high-resolution spectra of RHB candidates with the Robert 
G. Tull Cross-Dispersed Echelle spectrograph (\citealt{tul95}) of the 
2.7m Harlan J. Smith Telescope at McDonald Observatory. 
The instrumental setup, including a 1\farcs2 width entrance slit, 
yielded a spectral resolving power of 
R~$\equiv$~$\lambda/\Delta\lambda$~$\thickapprox$~60000.
Within the wavelength coverage of  $\lambda$$\lambda$ 3400-10900, we 
observed 63 spectral orders. 
The gaps between the orders reduce towards blue orders, enabling continuous
spectral coverage for $\lambda$~$\lesssim$~5900~\AA. 
The data were obtained during five
observing runs corresponding to 22 nights in total. 

Based on the selection criteria discussed in \S\ref{select}, we were
able to observe 129 candidate RHB stars with adequate signal-to-noise.
Not all of the stars survived scrutiny after data reduction.
Some of these stars were of interest because they were being observed 
with high spectral resolution for the first time. 
Inspection of the data revealed that 20 candidates either were
rapidly rotating ($V\sin{i}\geq$ 30) or were double-lined spectroscopic
binaries. These were discarded from the final sample because 
in both cases the spectral features become too blended and do not allow us to 
extract meaningful abundances for this project.
In the end we also had to eliminate 33 more candidates during 
the model atmosphere analyses because we were not be able to obtain 
reasonable atmospheric parameters from spectroscopic criteria alone 
(see \S\ref{params} for further discussion).
The remaining sample of 76 candidates were studied in detail.
In Table~\ref{tab-1} we present basic data for these program stars.

The data reductions were performed with the IRAF\footnote{The Image 
Reduction and Analysis Facility, a general purpose software package for 
astronomical data, is written and supported by the IRAF programming group 
of the National Optical Astronomy Observatory (NOAO) in Tucson, AZ, USA.} 
software package in the classical manner:  bias subtraction,
flat-field division, and scattered light removal followed by extraction 
of the spectral orders. 
We took at least two exposures per star and combined them in order to 
filter out cosmic rays events.
Fast-rotating hot stars were observed each night at appropriate air-masses,
and used to remove telluric features from the spectra of the program stars. 
Removal of the telluric lines was done with the IRAF routine 
$\it{telluric}$, following continuum normalization. 
ThAr lamp exposures taken at the beginning and end of each night were used 
for the wavelength calibration, again done with standard IRAF tasks.

Since some of the RHB candidates have reported parallaxes but 
no radial velocities (RVs), we also observed a few IAU RV standards.
We used these standards to measure the RVs of our stars via the
cross-correlating technique provided in $\it{fxcor}$ (\citealt{fitz93}) 
task in IRAF. 
The mean RVs of the relevant stars are listed in Table~\ref{tab-1}.
Typical errors for the RV measurements were around $\pm$0.3-0.6 \kmsec.
We also checked our measurements by re-measuring the RVs of a 
few of our program stars that have previously published values.
Our RVs agree with the literature values to within the joint 
error estimates.

We estimated signal-to-noise ($S/N$) ratios of the spectra 
at three different ``continuum'' wavelength regions near 4500, 5500 
and 6500~\AA\ in which we detect no obvious absorption features.
We give examples of the derived $S/N$ for three program stars with 
different visual magnitudes in Table~\ref{tab-ston}.

\section{MODEL ATMOSPHERE PARAMETERS\label{params}}

We have used both equivalent widths ($EW$s) and spectrum synthesis 
for the abundance analysis.
The line list for model atmospheric parameter determination was 
generated by choosing clean lines with laboratory $gf$-values between 
$\lambda$5500-6500~\AA.
This spectral region contains many relatively unblended Fe-peak species
transitions and only very weak CN molecular contamination in our
program stars.
In selecting the transitions we consulted the solar line compendium
of \cite{moo66} and the \cite{gri68} Arcturus spectral atlas.
We measured $EW$s using the SPECTRE\footnote{
An interactive spectrum measurement package, available at
http://www.as.utexas.edu/$\sim$chris/spectre.html} 
code (\citealt{fit87}).
SPECTRE employs a semi-automated routine that fits Voigt and Gaussian 
line profiles to the observed spectral lines.

\begin{figure}
\epsscale{.90}
\plotone{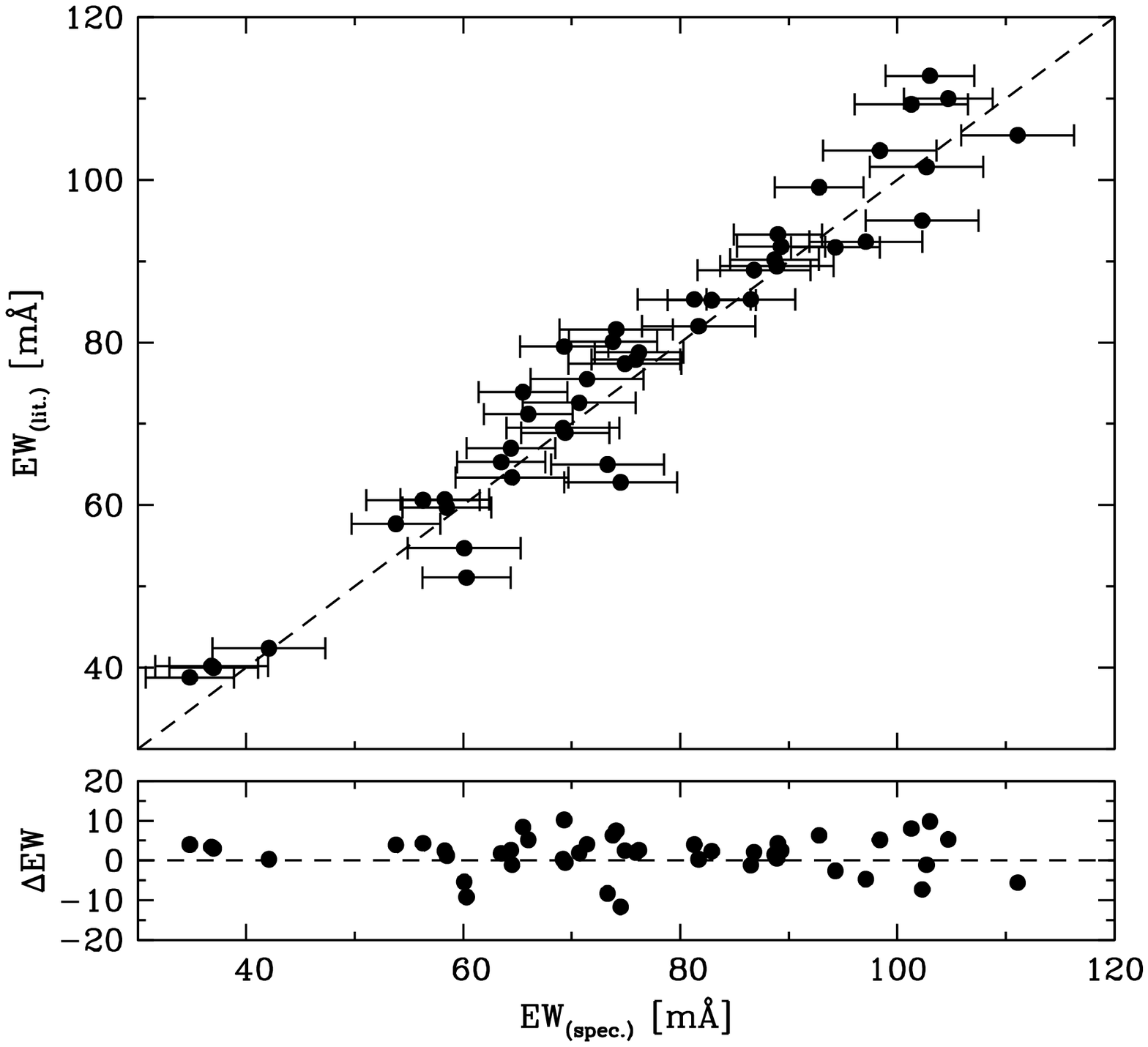}
\caption{\footnotesize 
         Top panel: comparison of our EWs with previous studies for 
         HIP~13339 and HIP 71837 (\citealt{tak05}). 
         Bottom panel: differences in EWs, defined as 
         $\Delta EW$~$\equiv$ $EW_{literature}$ $-$ $EW_{this study}$.
          Error bars represent the $\sigma$~$\simeq$~5~m\AA\ of $\Delta EW$.
\label{EW} \footnotesize
}
\end{figure}

In Figure~\ref{EW} we compare our EW measurements with the 
values given by \cite{tak05} for HIP 13339 and HIP 71837. 
The EW scales are in reasonable agreement: defining 
$\Delta$EW~$\equiv$ EW$_{literature}$ $-$ EW$_{this study}$,
we found $<$$\Delta$EW$>$~=~$+$1.6~m\AA,
with $\sigma$~$\simeq$~5~m\AA.

Stellar atmosphere models required by the analysis were from the 
\cite{cast97}, \cite{cas03} grid\footnote{
Available at: http://kurucz.harvard.edu/grids.html}
of model atmospheres computed with opacity distribution functions and
without convective overshooting.
We interpolated in this grid with software developed by
Andy McWilliam and Inese Ivans.
Then we used an automated version of the spectral line analysis 
and synthetic spectrum code MOOG (\citealt{sne73})\footnote{
available athttp://www.as.utexas.edu/$\sim$chris/moog.html} 
to determine the abundances of our program stars. 
MOOG performs analyses using one-dimensional local thermodynamic
equilibrium (LTE) equations for plane parallel atmospheres.
In the automated version (described in more detail by \citealt{hol11}
and Roederer et~al., in preparation) all of the steps in model atmosphere
derivation and many parts in the relative abundance analysis are done
iteratively by the code without human intervention.

We used \ion{Fe}{1} and \ion{Fe}{2} abundances derived with trial
model stellar atmospheres to determine the fundamental stellar parameters
of effective temperature \teff, surface gravity, \logg, microturbulent
velocity \vmicro, and metallicity [Fe/H]\footnote{
For elements A and B, 
[A/B] = log $(N_{A}/N_{B})_{\star}$ -- log $(N_{A}/N_{B})_{\sun}$ 
and log $\epsilon$(A) = log $(N_{A}/N_{H})$ + 12.0}.
Final values of \teff\ were estimated by requiring that \ion{Fe}{1} 
abundances show no trend with excitation potential $\chi$ beyond
internal line-to-line scatter uncertainties.
A similar approach yielded estimates of \vmicro, which is related to 
small-scale turbulent motion: this parameter was varied until we
obtained no apparent \ion{Fe}{1} abundance trend with reduced 
width $RW$ ($\equiv$~$EW/\lambda$).

As reported in previous studies, non-local thermodynamical 
equilibrium (NLTE) mechanism almost has no effect on 
\ion{Fe}{2} lines and has a small effect on \ion{Fe}{1} lines 
for solar-type metallicities and mildly metal-poor stars. As given by
\cite{mash11} for a temperature range of 4600$-$6500~K, the departure from 
LTE is not more than 0.1 dex in \ion{Fe}{1} abundances. 
This value is well within the uncertainty limits of our \ion{Fe}{1}
abundances (see \S\ref{errors}).  
Since we have mostly solar-metallicity and mildly metal poor stars in
our program, the corrections for NLTE effects were not applied.

Surface gravities were calculated by demanding that \ion{Fe}{1} and 
\ion{Fe}{2} lines yield the same mean abundances to within the 1$\sigma$
internal scatter uncertainties of both species.
Finally, we iterated on the [Fe/H] metallicities until the values
assumed in model creation were consistent with those implied by
the line abundance averages.
Derived model parameters of our stars are given in Table~\ref{tab-model}.

We also derived solar abundances (Table~\ref{tab-sun}) applying the 
same procedure to all the species investigated here. 
The high-resolution solar data were obtained from 
the electronic version of the solar center-of-disk spectrum of
\cite{del73}\footnote{
Available at the Bass2000 web site, http://bass2000.obspm.fr/}.
These abundances were used for differential determination of the 
stellar abundances. 
These differential abundances should give more accurate ``internal'' 
results than absolute abundances since many of the systematic errors 
can be nearly compensated in comparing our stars with the Sun.
Our internal solar abundances generally agree with those recommended
by \cite{asp09}, to within the mutual uncertainties.
Nitrogen is the sole exception with a substantial difference, 
$\sim$+0.15 dex, which potentially could lead to a small offset in our 
[N/Fe] results compared to the previous studies that we will consider
in \S\ref{pcapture}.
Our adopted CN oscillator strengths from the Kurucz database may be the 
major factor contributing to this offset.
Exploration of this issue in detail is beyond the scope of our paper.

\subsection{Uncertainties in Parameters\label{errors}}

The internal uncertainties in atmospheric parameters 
\teff, \logg, and \vmicro\ were estimated by running a series of trial 
analyses on the spectral data of BD+27~2057, HD~84686 and HIP~98587. 
For temperature uncertainties, we varied the assumed \teff\ in steps 
of 50~K and kept the other parameters fixed during the analysis. 
The temperature was changed until the mean abundance difference 
between low and high excitation \ion{Fe}{1} lines exceeded the 
$\pm$1$\sigma$ scatter of individual line abundances derived with the 
optimal \teff, in other words, until \ion{Fe}{1} abundances showed an 
obviously unacceptable trend with excitation potential.
This method yielded an average uncertainty of $\sim$150~K for the \teff.

\begin{figure}
\epsscale{.90}
\plotone{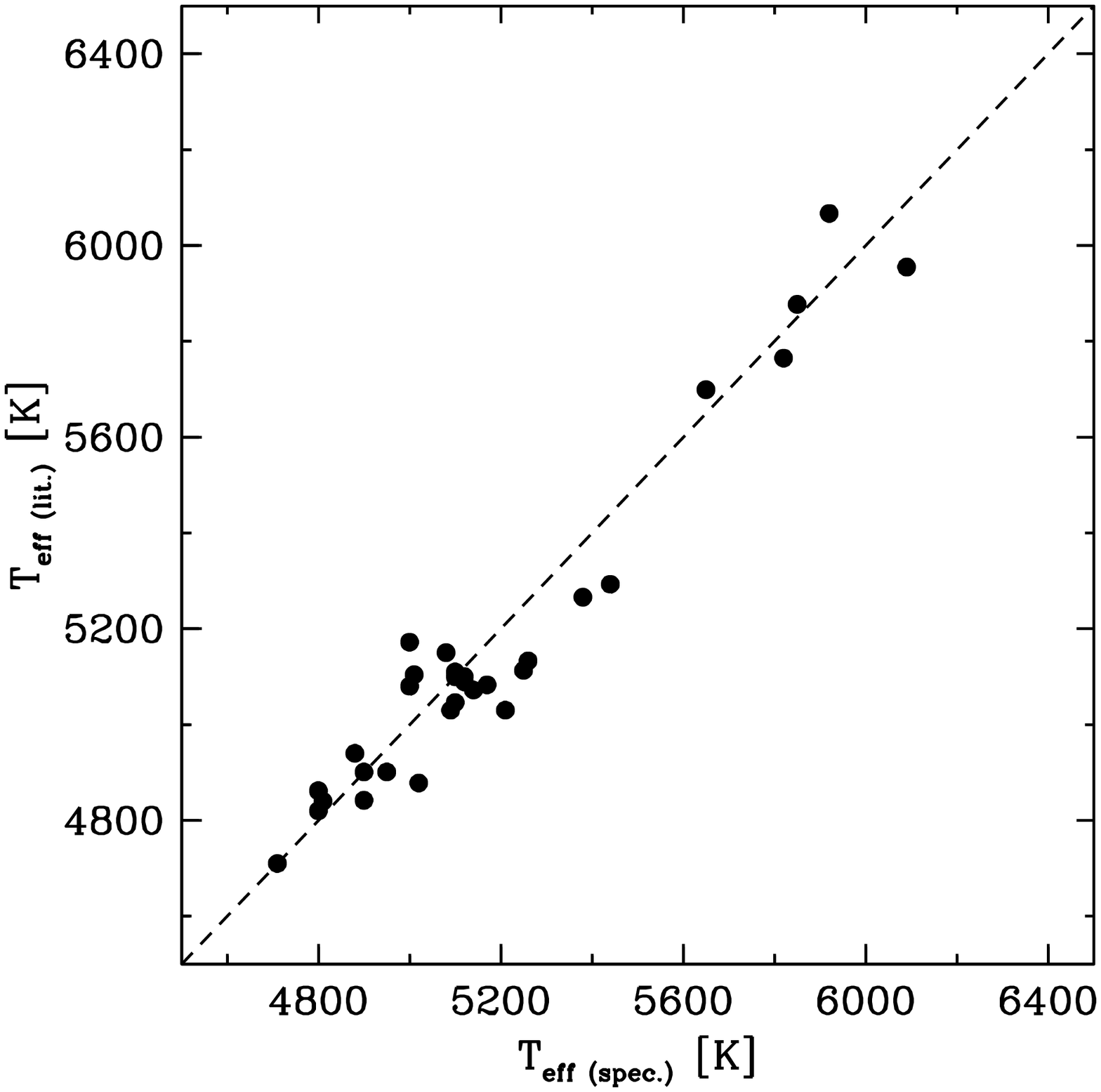}
\caption{\footnotesize
         Comparison of our derived spectroscopic \teff\ values with 
         those gathered from heterogeneous literature (filled circles):
         \cite{cott86}, \cite{mcw90}, \cite{rya95}, \cite{mal98}, 
         \cite{che00}, \cite{tau01}, \cite{fra04}, \cite{nor04}, 
         \cite{luc07}, \cite{rob07}, \cite{sou08}, \cite{tak08},
         \cite{omi09}, and \cite{liu10}.
\label{Tefflit}
}
\end{figure}

To estimate the external uncertainty in \teff\ we have 
searched the literature for previous high-resolution studies of
our program stars.
We found no large samples in common with our stars, but from the \teff\
values reported in various publications we have generated 
Figure~\ref{Tefflit}. 
The heterogeneity of the literature data does not justify a detailed
statistical treatment, but it is clear that our \teff\ values
track those in previous publications, with a scatter of $\sim$120~K.
We adopt an overall \teff\ uncertainty of $\pm$150~K.

Our \teff\ values for a few stars, however, deviate significantly
from previous estimates.
Since temperature estimates are critical to assessment of RHB or RC
status for our stars, we turned to the spectroscopic ``Line Depth Ratio"
(LDR) method of \teff\ determinations.
\cite{gray01} showed that simple LDRs of several line pairs in the
6200~\AA\ spectral region (most often a ratio formed by
comparing the central depths of a \ion{V}{1} and an \ion{Fe}{1} line)
Their paper maps the LDRs to stellar $B-V$ values and those colors
in turn to \teff\ estimates.
LDRs are excellent temperature indicators especially 
for giants with spectral types between G3 and K3 (see also Figure~7 of
\citealt{gray01}). 
This method produces effective temperatures well within the uncertainty 
levels of 150~K and gives us an independent spectroscopic check on
our spectroscopic \teff\ estimates from \ion{Fe}{1} lines.

We measured LDRs for our whole sample.
We mainly made use of the line ratio pairs recommended by 
\cite{gray01}\footnote{The line ratio pairs used for LDR temperatures: \ion{V}{1}(6224.5~\AA)/\ion{Ni}{1}(6223.9~\AA) , 
\ion{V}{1}(6233.2~\AA)/\ion{Fe}{1}(6232.6~\AA), \ion{V}{1}(6242.8~\AA)/\ion{Si}{1}(6243.8~\AA),
\ion{V}{1}(6251.8~\AA)/\ion{Fe}{1}(6252.5~\AA), \ion{V}{1}(6256.8~\AA)/\ion{Fe}{1}(6255.9~\AA)}
and derived an average \teff(LDR) for the stars. 
These line pairs compare the central depths of a very \teff-sensitive 
transition (almost always a low-excitation\ion{V}{1} line) with
usually a high-excitation transition of another Fe-group species.
We especially made use of the 6224.5~\AA/6223.9~\AA\ pair that
compares a \ion{V}{1} line with $\chi$~=~0.29~eV to a \ion{Ni}{1}
line with $\chi$~=4.10~eV.
Their LDR was the most useful because it could be reliably measured
throughout most of the \teff\ domain of our sample.
However, the depths of \ion{V}{1} lines weaken rapidly
with increasing temperature ($\geq$ 5500~K).
For example, for the five \ion{V}{1} lines recommended by \cite{gray01}
for LDR studies, $<EW>$~$\simeq$~6~m\AA\ in the solar spectrum
(\citealt{moo66}).
Lines with such small $EW$s are near the reliable detection/measurement
limit of our spectra.
Therefore for warmer stars of our sample we also measured some LDRs 
proposed by \cite{stras00}.

\begin{figure}
\epsscale{.90}
\plotone{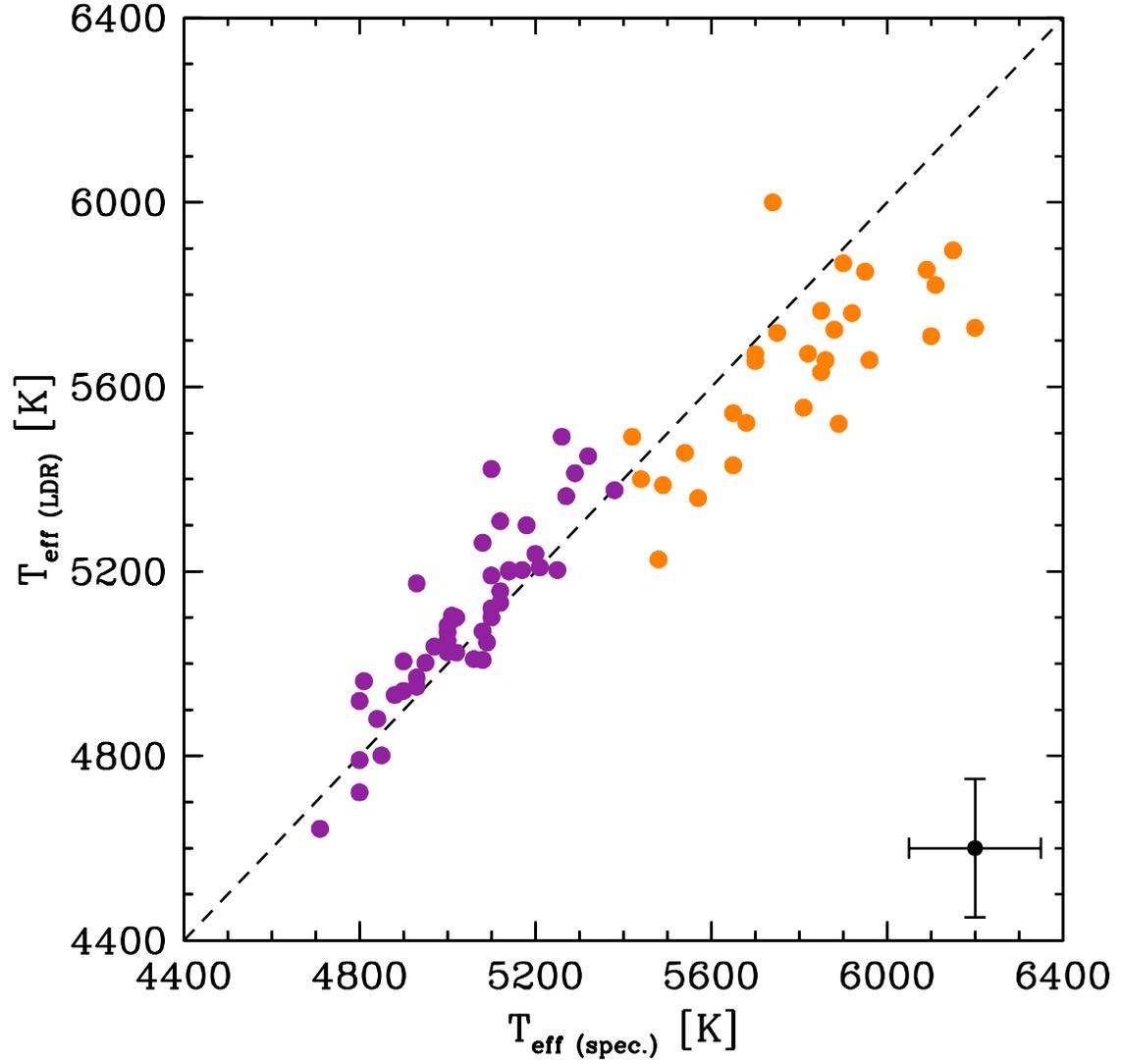}
\caption{\footnotesize
         Comparison of our spectroscopic \teff\ with the ones obtained
         using LDR method.
         A break in the relation is seen at \teff~$<$ 5500~K,
         which we call attention to by using different symbol colors for 
         stars warmer and cooler than this temperature; see text for 
         discussion of this issue.
\label{LDR}
}
\end{figure}

A comparison of our spectroscopic and LDR temperatures 
are given in Figure~\ref{LDR}. 
The error bar depicted in this figure corresponds to the 
standard deviation of the mean difference between these two 
temperatures, $\sigma$~$\simeq$~$\pm$150~K. 
A change in the correlation is apparent at \teff~$<$~5500~K,
which we have highlighted by using different colors for the symbols
of stars warmer and cooler than this \teff. 
Although we see a small systematic bias towards higher
temperatures for \teff\ $<$ 5500~K (Figure~\ref{LDR}, purple points),
the standard deviation of these points is around 90 K and well within the 
uncertainty limits. We suggest that this systematic deviation may 
both arise from the method we adopt from \cite{gray01}, which first 
involves a calibration of LDR against B$-$V color indices,
then converts these into temperatures, and LDRs that we obtained 
with different instrumentation than those of \cite{gray01}.
As mentioned above, this temperature is near the reliability limit 
for the LDR method using these transitions.
On the other hand, for \teff\ $<$ 5500~K, which 
proves to be the more important domain for identifying true
RHB targets, these two \teff\ estimators are considerably in good 
agreement. 
This result suggests that for stars when clashes occur between
photometric and spectroscopic temperatures, the photometric
values are probably not reliable for \teff~$<$~5500~K.
In general, LDR method supports our claimed temperatures
for the stars, including the ones which our spectroscopic 
values clash with literature estimates.
  
We determined internal uncertainties for \logg\ and \vmicro\
through repeated trials with variations in these quantities. 
The typical average uncertainties were estimated as 0.16~dex and 
0.2~\kmsec, respectively. 
Comparison of our \logg\ results with previously reported ones resulted in 
an external uncertainty level of $\approx$0.25~dex.
By taking into account both internal and external uncertainty levels, 
we adopt an average uncertainty for \logg\ of $\approx$$\pm$0.3 dex.

\begin{figure}
\epsscale{.90}
\plotone{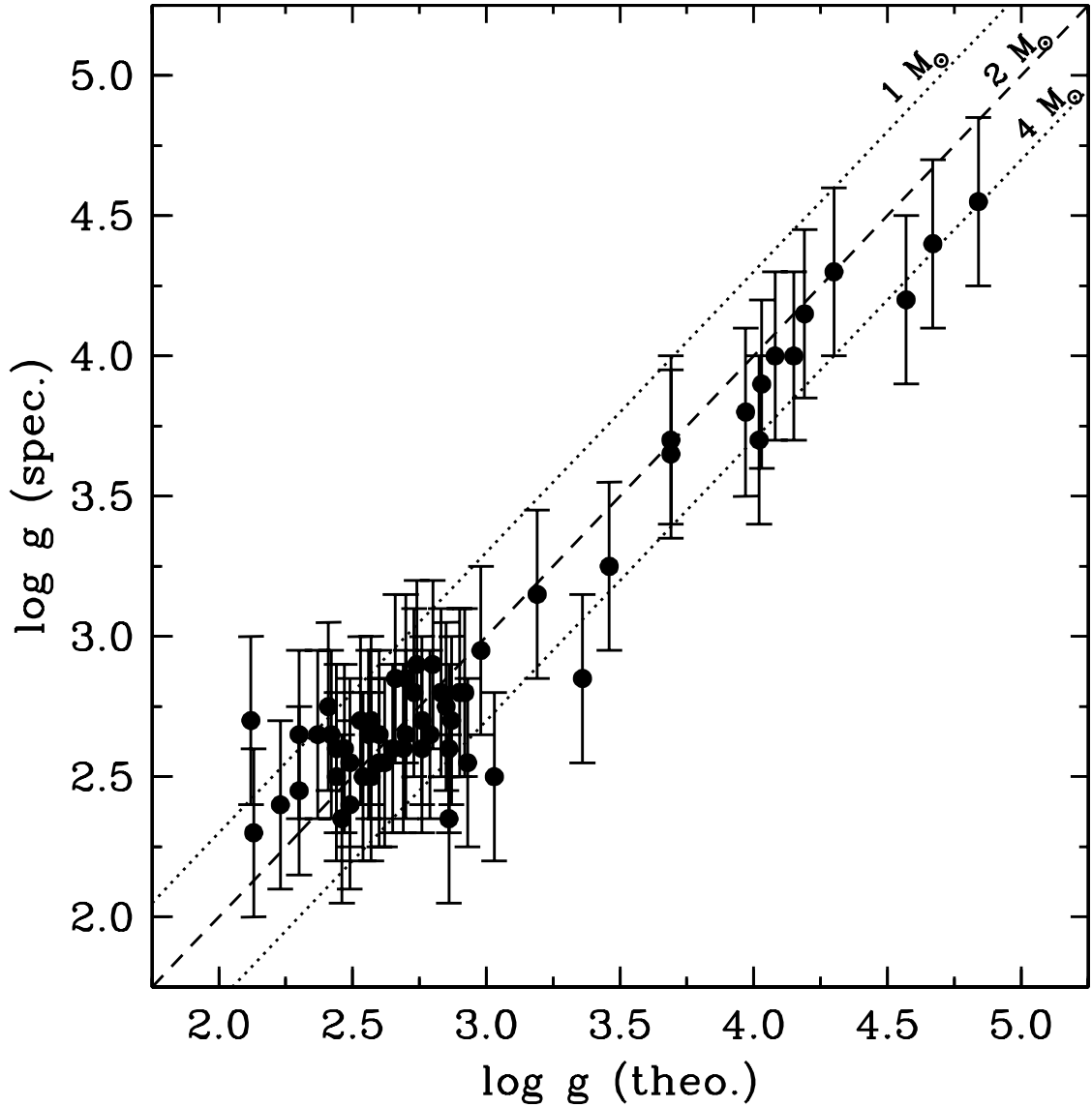}
\caption{\footnotesize
         Comparison of our spectroscopic \logg\ values with theoretical 
         ones. \logg(theo.) are calculated for M =1, 2 and 4 M$_{\sun}$ (dashed and dotted lines).
         The uncertainty for our spectral \logg\ is $\pm$0.3 dex. We do not have errors
         on the theoretical values since those are adopted masses.
\label{logg}
}
\end{figure}

We also calculated ``physical'' gravities and compared them 
with spectroscopic gravities in order to possibly gain insight on 
the masses of our stars.
We use the following standard equation for physical gravities
\begin{displaymath}
\logg_{\star} = 0.4 (M_{\rm V\star} + BC - M_{\rm Bol\sun}) + \logg_{\sun} + 
4 {\rm log} (\frac{\teff_{\star}}{\teff_{\sun}}) + {\rm log} (\frac{{\it m}_{\star}}{{\it m}_{\sun}}).
\end{displaymath}
For the Sun, we adopted \teff = 5780 K, \logg = 4.44 dex, and 
$M_{\rm Bol\sun}$ = 4.75 mag.
Bolometric corrections were calculated using the relation given 
by \cite{alo99}.
A comparison of physical gravities, \logg(theo.), with spectroscopic 
gravities, \logg(spec.), is given in Figure~\ref{logg}. 
A significant uncertainty in a physical gravity calculation is 
the assumed mass. 
Therefore in the figure we give three \logg(theo.) values calculated 
for the masses of 1, 2 and 4 M$_{\sun}$. 
They are shown with dashed and dotted lines. 
The best average agreement between these two \logg\ scales 
for significantly evolved stars (\logg~$\lesssim$~3.0) is for 
M~$\simeq$ 2 M$_{\sun}$ and for MS/SG stars it is M~$\simeq$ 2.5 M$_{\sun}$.
The joint uncertainties are too large to draw conclusions about individual
stars, but does suggest that our sample is not dominated by high 
mass stars.

In order to derive the effect of these uncertainties on abundance 
determinations, we did multiple analyses by changing \teff, \logg, [Fe/H], 
and \vmicro\ within their uncertainty limits.
The effective temperature uncertainties create an uncertainty 
around 0.15-0.16 dex in \ion{Fe}{1} and the uncertainties in \logg~ 
make the \ion{Fe}{2} abundances change in less than 0.1 dex.
The other elements' abundances vary in between 0.1-0.16 dex due to
\teff~uncertainties. In general, the uncertainties in surface gravity and 
microturbulent velocity result in abundance uncertainties 
less than $\simeq$0.1 dex.

Finally we comment on the uncertainties in our derived \ciso~ ratios.
These are nearly insensitive to model atmosphere uncertainties,
since \iso{12}{CN} and \iso{13}{CN} are nearly identical molecules.
We estimated the \ciso\ uncertainties by fitting synthetic spectrum
to the observed \iso{12}{CN} and \iso{13}{CN} features with various 
isotopic ratios to estimate their maximum and minimum probable values.
The error estimates for the \ciso~ratios are given in Table~\ref{tab-CN}.

\subsection{Photometric Temperatures and Reddening\label{teffcomp}}

Since we observed almost half of our program stars at high spectral
resolution for the first time, our sample lacked a consistent set of
reported atmospheric parameters. 
This made it difficult to have meaningful temperature estimates 
to start our analyses, so we relied exclusively on our \ion{Fe}{1}
line analyses to derive \teff\ values, 
augmenting these with LDR estimates.
Armed with these results we now compare spectroscopic and photometric
temperatures for the stars in common.
We concentrate on the $V-K$ color because it is one of the best photometric
temperature indicators, having almost no dependence on metallicity and gravity 
(e.g. \citealt{alo99}, \citealt{ram05}).
The $V$ magnitudes that we employ are on the standard Johnson photometric
system; the $K_{s}$ magnitudes are from the Two Micron All Sky Survey.
We adopt the ($V-K_{s}$)-\teff\ calibration equation from \cite{ram05}.

Before calculating the photometric \teff\ values, we first applied 
interstellar reddening corrections to the $V-K_{s}$ colors. 
A recent survey of the regions within 300~pc of the Solar System showed 
that the ``Local Bubble" (\citealt{lal03}) has a shape of an irregular 
lacuna, which extends $\sim$60~pc towards the Galactic center, 80$-$150~pc 
towards the outermost edge of the Galaxy and $\sim$200~pc above and 
below the Galactic plane. 
But ISM dust extinction is patchy and varies considerably toward
different sightlines, and many of our stars have poorly-constrained distances.
Therefore we estimated their reddening in three different ways.
In all cases, we adopted a typical Local Bubble radius of 75~pc and 
assumed no reddening for stars estimated to be within that
distance (e.g. \citealt{hen00}).

For our first reddening estimate, we have applied an isotropic 
reddening correction for the stars have distances 75~$<$~$d$~$\leq$ 300~pc, 
adopting $A_{v}$~=~0.8 mag~kpc$^{-1}$ (\citeauthor{hen00}).  
The extinction values of eight stars with distances $d$~$>$~300~pc 
were obtained from the NASA/IPAC Extragalactic Database (NED)\footnote{
http://ned.ipac.caltech.edu/}
extinction calculator. 
Zero reddening was assumed for the stars with high Galactic latitudes, 
$|b|$~$\geq$~50$^o$ (e.g, \citealt{san72}).
For the second and third reddening estimates, we employed the methods 
described by \cite{che98} and \cite{hak97}. 
Our computations with the latter method used Hakkila's EXTINCT 
code\footnote{
Available at http://ascl.net/extinct.html}
which takes into account various interstellar extinction correction 
methods from several studies (see \citeauthor{hak97} for more details).

\begin{figure}
\epsscale{0.6}
\plotone{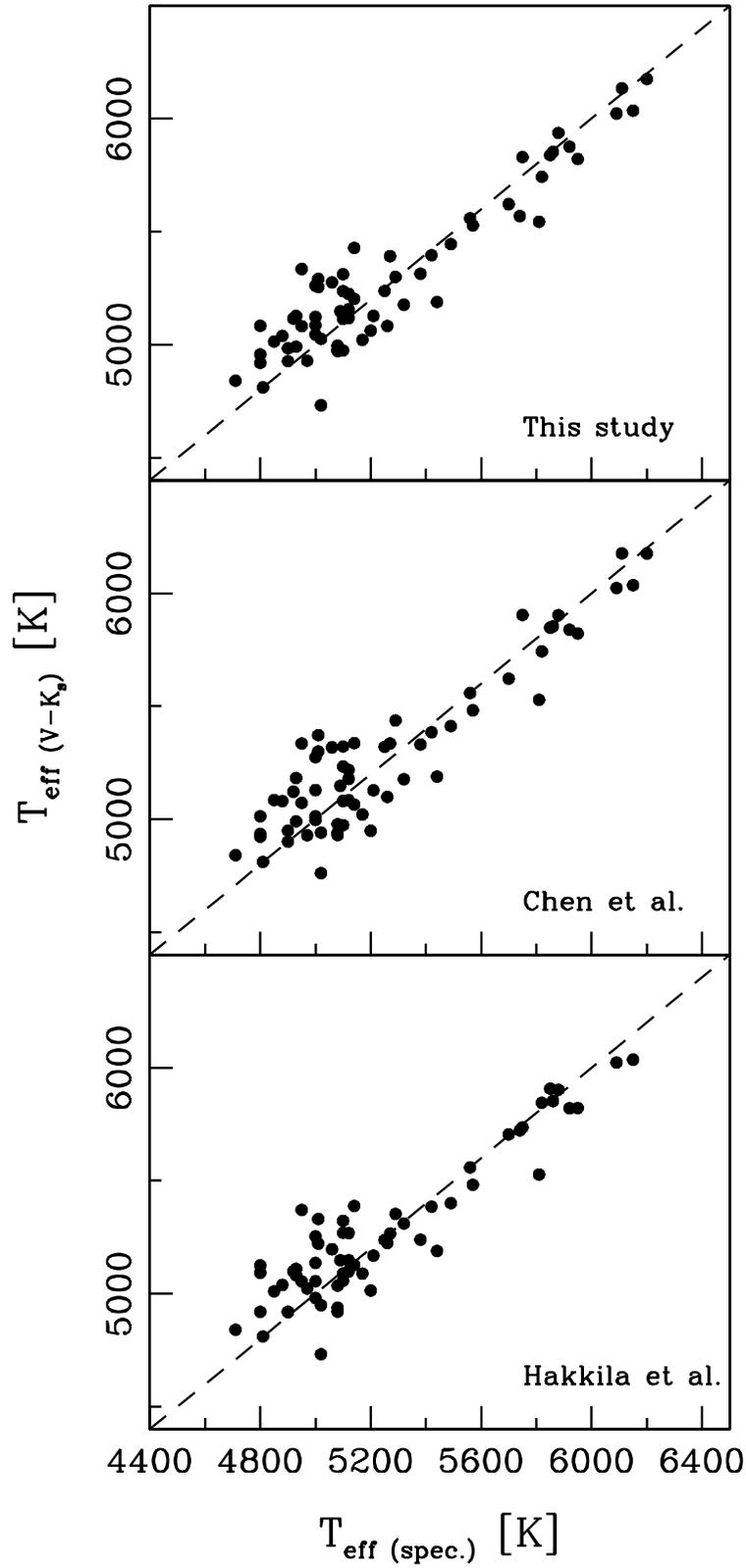}
\caption{\footnotesize
         Comparison of spectroscopic \teff\ (spec.) with photometric \teff\ (V--K$_{s}$),
         adopting the three different reddening estimates 
         discussed in \S\ref{teffcomp}.
\label{fig1}
}
\end{figure}

We applied each of these reddening estimates to the observed 
$V-K$ colors, computed photometric temperatures from the reddening-corrected 
$V-K$ colors using the \cite{ram05} formula, and correlated them with our
spectroscopic temperatures as shown in Figure~\ref{fig1}.
Inspection of this figure clearly suggests that the two \teff\
scales are well correlated, and that various reddening assumptions
produce little substantive variations in the temperature correlations.
Therefore, we have adopted our E(\bv) values for reddening for the
entire sample.

\section{ABUNDANCE DETERMINATIONS\label{abunds}}

Using the model atmospheres given in Table~\ref{tab-model}, we 
determined abundances of several elements in our program stars.
We report these as relative abundance ratios [X/H] in Table~\ref{tab-abun}. 
Whenever possible, we computed the abundances from line EWs.
For complex transitions, those that have significant hyperfine and
isotopic substructure, and those that have significant line blending
issues, we resorted to synthetic/observed spectrum matches to determine
the abundances.
We were especially interested in three element groups that could help 
constrain the Galactic population memberships of our stars and verify
their evolutionary states.
In the following subsections we consider in turn the $\alpha$,
the neutron-capture, and the proton-capture element groups.

\subsection{$\alpha$-Elements: Silicon and Calcium\label{alpha}}

The abundances of $\alpha$-elements exhibit the same behavior 
in giant and dwarf stars of solar metallicity ([$\alpha$/Fe]~$\sim$~0;
\citealt{soubiran05}, \citealt{mishenina06}). 
In thick disk main-sequence and subgiant stars, the $\alpha$ elements
typically become overabundant
as metallicity decreases, reaching 
[$\alpha$/Fe]~$\sim$~$+$0.3 at [Fe/H]~$\sim$~$-$1 (e.g. \citealt{tau01}, 
\citealt{tau01}, \citealt{reddy06}).
The temperatures and gravities of these prior large samples are
5000~K~$\lesssim$ \teff~$\lesssim$ 6500~K and 
3.4~$\lesssim$ \logg~$\lesssim$ 4.7.
Less systematic abundance trend information is available for thin/thick
disk giant stars, which in our case cover the approximate parameter range
4800~K~$\lesssim$ \teff~$\lesssim$ 5600~K and 
2.2~$\lesssim$ \logg~$\lesssim$ 4.0.

The easily-observable $\alpha$ elements are Mg, Si, and Ca.
Often Ti is grouped with the other $\alpha$'s due to its 
similar abundance behavior with metallicity. 
However, Ti is not a pure $\alpha$ element because its dominant isotope 
is $^{48}$Ti$_{22}$, which is not an even multiple of $\alpha$ particles.
Here we concentrated on Si and Ca abundances, since they have large numbers
of transitions with a range of line strengths in the yellow-red 
spectral region (most \ion{Mg}{1} lines are very strong in our G-K giant 
stars).
We generally used about 9 \ion{Si}{1} and 15 \ion{Ca}{1} lines, and
derived abundances from their EWs.

\begin{figure}
\epsscale{1.00}
\plotone{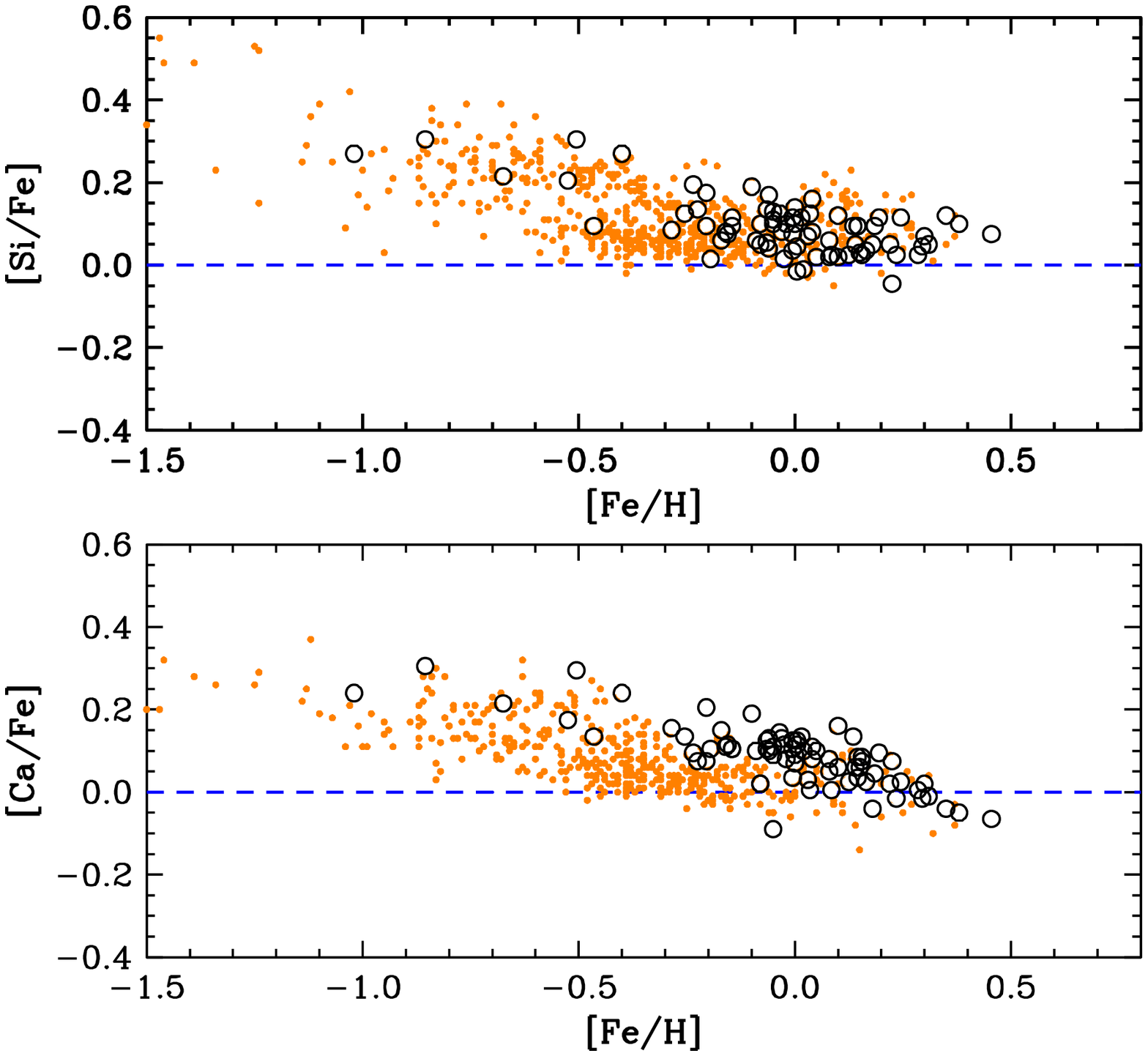}
\caption{\footnotesize
         Relative Si and Ca abundances in our program stars 
         (black open circles) and in stars in the literature 
         (orange dots).
         All results are based on lines of \ion{Si}{1} and \ion{Ca}{1} species.
         The data from the literature are adopted without adjustment
         from \cite{tau01}, \cite{bensby03}, \cite{reddy03,reddy06} 
         and \cite{mishenina06}.
\label{SiCaFe_litC} 
}
\end{figure}

We compare our Si and Ca abundances with published values in 
Figure~\ref{SiCaFe_litC}. 
The literature data are taken from several studies that are given in 
the figure caption: \cite{tau01}, \cite{reddy03}, \cite{bensby03}, 
\cite{reddy06} and \cite{mishenina06}. 
Our sample has a wide metallicity range 
$-$1.0~$\leq$~[Fe/H]~$<$~$+$0.5 and it is clear that the 
$\alpha$ elements behave similarly in our cooler
giants as they do in warmer main sequence stars and subgiants.
Even though slight scaling differences ($<$ 0.1 dex) naturally result 
from different solar abundances, different line choices, and different
oscillator strengths adopted here compared with previous studies,
our results are generally in good agreement with the literature.
At a given [Fe/H] metallicity our mean [Ca/Fe] values are typically within
0.05~dex of literature values, and [Si/Fe] are within 0.01~dex.

\subsection{Neutron-Capture Elements\label{ncapture}}

Among thin and thick disk stars, those elements whose solar-system
origin is due chiefly to ``slow'' neutron bombardment reactions
(the $s$-process; e.g. Sr, Y, Zr, Ba, La, and Ce) generally
exhibit their solar abundance ratios ([X/Fe]~$\sim$~0) throughout
metallicity regime $-$1.0~$\lesssim$ [Fe/H]~$\lesssim$ $+$0.2.
But those elements that are products of ``rapid''-blast neutron
capture events (the $r$-process; Eu, Gd, Dy) tend to increase 
in relative abundance with decreasing metallicity.
The most easily-observed $r$-process element is Eu, and its mean
abundances reaches [Eu/Fe]~$\sim$~+0.4 at [Fe/H]~$\sim$~$-$1.
A good summary of the disk $s$- and $r$-process abundance trends
can be seen in Figure~17 of \cite{reddy06}.
Additionally, \cite{sim04} has shown that a distinct kinematic
signature in disk neutron-capture elements: stars with larger
space motions have often much lower $s$-process abundances than $r$-process
ones (see their Figure~12).

There are not a lot of useful neutron-capture element (Z~$>$~30) 
transitions in the yellow-red spectral region.
We considered just La (75\% $s$-process origin in solar-system 
material, e.g., \citealt{sneden08} and references therein), and
Eu (97\% $r$-process).

\textbf{\ion{La}{2}:} transitions at 6262.2 and 6390.5~\AA\ can be
detected in nearly all of our stars.
These lines have well-determined transition probabilities and 
hyperfine structure parameters (\citealt{lawler01a}; see also 
\citealt{ivans06} for complete substructure line lists for these transitions).
The sole naturally-occurring isotope of this element is \iso{139}{La}.
Because of the complexity of these \ion{La}{2} transitions we used 
synthetic spectrum analyses to determine the abundances.
Notable contaminants to the La features are CN red-system lines,
but in most cases the CN strengths were small and did not materially
affect the derived abundances.

\textbf{\ion{Eu}{2}:} transitions at 6645.1 and 7217.5~\AA\ were used
in the Eu abundance analyses.
The transitions are very complex, because Eu has both hyperfine
and isotopic substructure.
There are two Eu stable isotopes, \iso{151}{Eu} (48\% of the total
solar-system Eu abundance) and \iso{153}{Eu} (52\%).
There is little difference in Eu isotopic ratios generated in 
$r$-process and $s$-process environments. 
Spectroscopic studies of \ion{Eu}{2} in extremely metal-poor, 
Eu-enhanced stars suggest that the \iso{151}{Eu} fraction is $\approx$50\% 
for $r$-rich stars (\citealt{sne02}, \citealt{aoki03a}, \citealt{roe08}) 
and $\approx$57\% for $s$-rich stars (\citealt{aoki03b}).
Therefore in our Eu syntheses we adopted the solar-system isotopic ratios.
The basic laboratory analyses of these lines were published
by \cite{lawler01b}; full hyperfine/isotopic substructure lists are
in \cite{ivans06}.

\begin{figure}
\epsscale{1.00}
\plotone{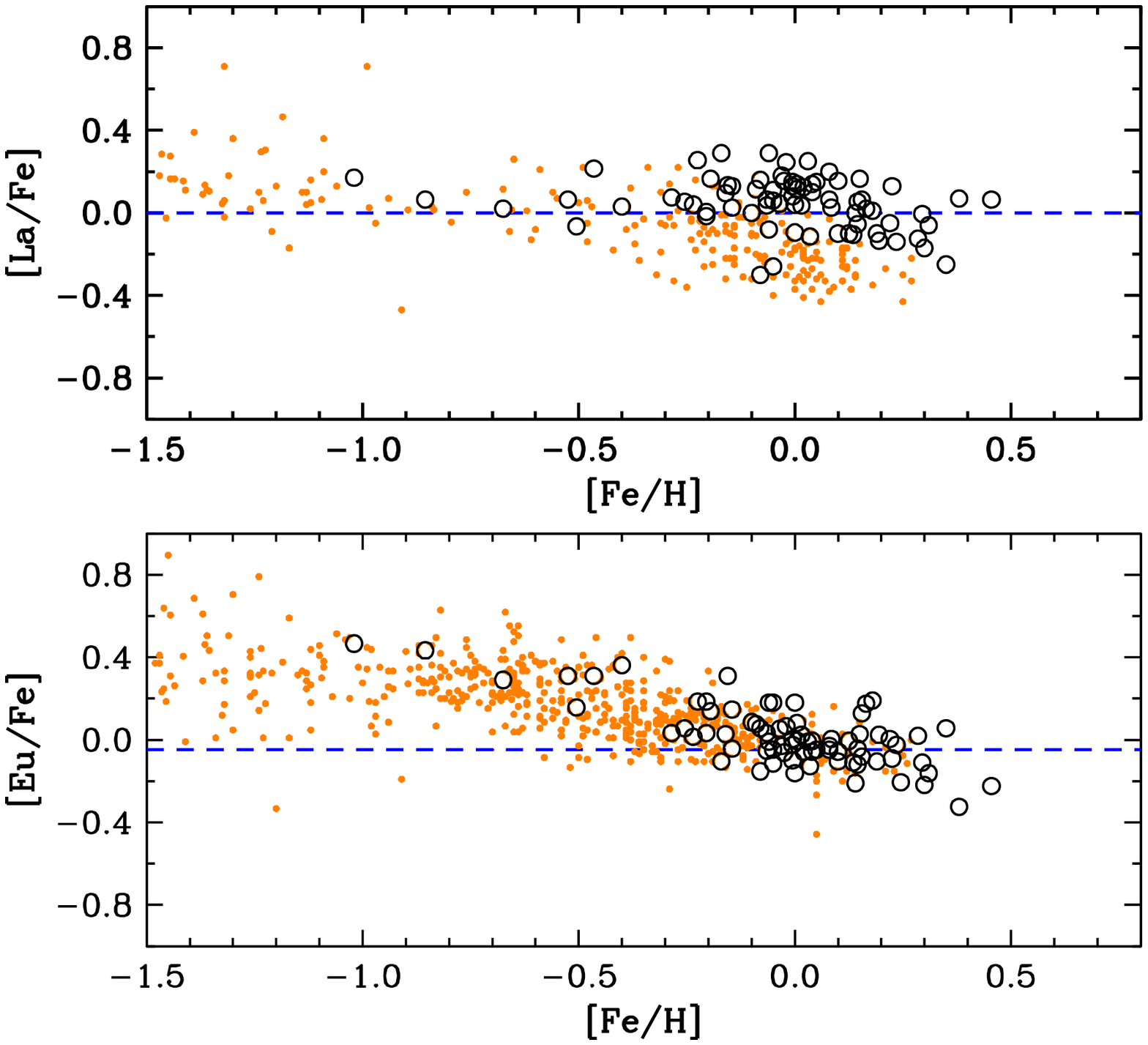}
\caption{\footnotesize
         Relative La and Eu abundances in our program stars
         (black open circles) and in stars in the literature
         (orange dots).
         All results are based on lines of \ion{La}{2} and \ion{Eu}{2} species.
         The data from the literature are adopted without adjustment
         from \cite{tau01}, \cite{bensby03}, \cite{venn04}, \cite{sim04}, \cite{reddy03,reddy06} and \cite{mishenina07}. 
\label{LaEuC}
}
\end{figure}

The abundance variations for \ion{La}{2} and \ion{Eu}{2} 
as a function of metallicity are given in Figure~\ref{LaEuC}. 
The observed trends (no apparent change in [La/Fe], increasing
[Eu/Fe] at lower metallicities) are in complete agreement
with previous results cited above.
The neutron-capture element abundances, combined with those of the
$\alpha$~elements clearly provide Galactic population indicators
for our stars.

\subsection{The Proton-Capture Abundances\label{pcapture}} 

Evolved Population~I solar-metallicity stars show evidence
of convective envelope mixing; their observed light elements Li, C, 
N, and O clearly have been altered via interior synthesis in 
$p-p$ and CNO cycles.
The original C abundance drops by a factor of about two, the N abundance
rises by comparable amounts, and the carbon isotopic ratio drops
to values usually between 15$-$30 (e.g. \citealt{lamb81}).
Mildly metal-poor, high-velocity stars (thick disk, labeled old disk in
early papers) of about a solar mass have lower \ciso\ ratios but
less evidence for depleted C and enhanced N 
(e.g., \citealt{cott86}).
Low mass giant stars of the Galactic halo often display more 
dramatic \iso{12}{C}$\rightarrow$\iso{13}{C} and C$\rightarrow$N
conversions (e.g., \citealt{grat00}).
Thus the CNO abundances of evolved stars can yield information on 
their internal evolutions and population memberships.

Since stars easily destroy Li in relatively low-temperature
proton-capture reactions, Li abundances or upper limit estimates in 
metal-rich stars provide additional information.
The presence of Li in the spectrum of an evolving star usually suggests 
that envelope convection has not yet developed to a point where 
dredge-down of original surface Li has effectively cleaned it from the 
star's envelope.  
The absence of Li can indicate (but not always) that other 
proton-capture products may have been dredged up to the surface.
This straightforward interpretation is complicated by the relatively 
rare phenomenon of Li-rich giants (e.g., \citealt{char00}, 
\citealt{kumar11} and references therein).
This increases the importance of searching for this element
in the spectrum of any evolved star.

Unfortunately, many atomic and molecular transitions that 
are CNOLi abundance indicators have significant detection and/or
analytical issues.
Here we describe the transitions that we used, and our 
atomic/molecular parameter choices.

\textbf{\ion{C}{1}:}  High excitation ($\chi$~$\gtrsim$~7.7~eV) lines
of this species are strong in the warmest candidate stars but weaken
with decreasing \teff.
We adopted the transition probabilities for these lines recommended
in the NIST Atomic Spectra Database (\citealt{ral11})\footnote{
Available: http://physics.nist.gov/asd3}.
However, the \ion{C}{1} lines decrease in strength so quickly with 
decreasing \teff, often only a couple of them could be used for
C abundances in our stars. 

\begin{figure}
\epsscale{1.00}
\plotone{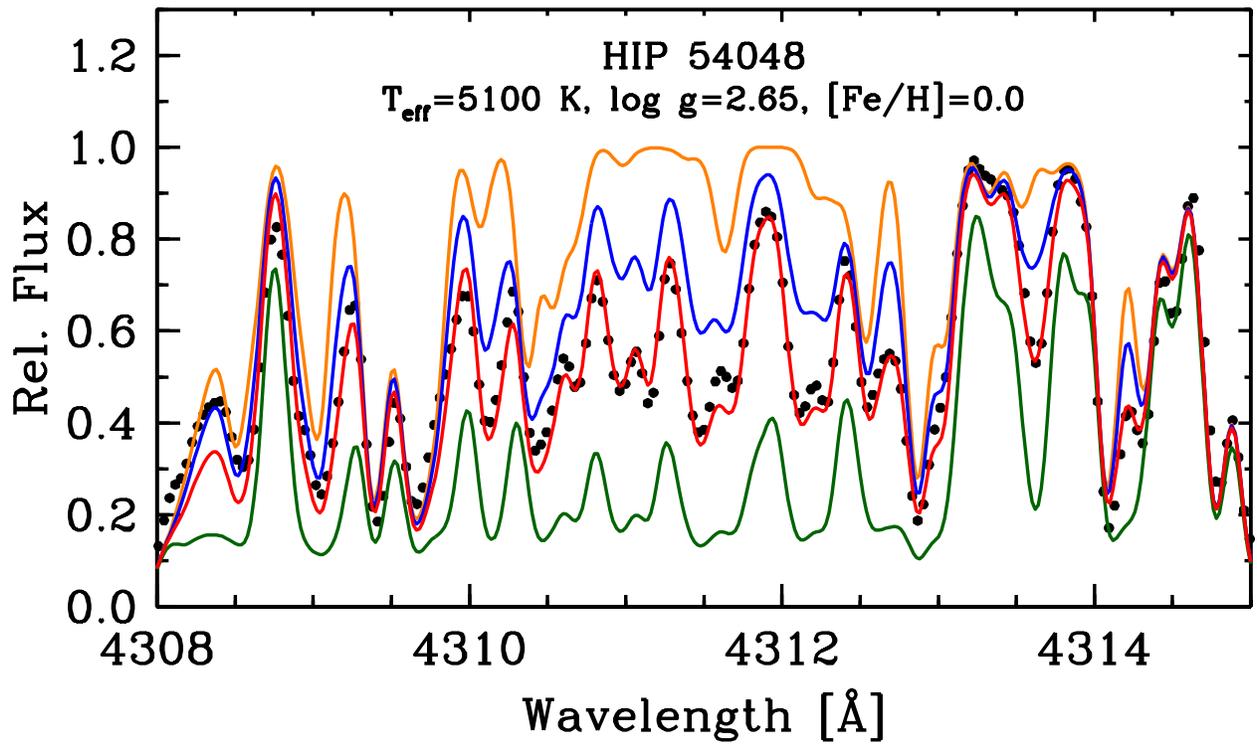}
\caption{\footnotesize
         Comparison of synthetic and observed spectrum of HIP 54048 
         for a part of the CH G-band region. Best fit is illustrated 
         by a red solid line for log~$\epsilon$(C)=7.79. Other fits represented
         by yellow, blue and green solid lines are given for
         log~$\epsilon$(C)=2.79, 7.34 and 8.69, respectively.
\label{HIP54048_CH} 
}
\end{figure}

\textbf{{CH G-Band:}} We also employed the CH $X^2\Pi-A^2\Delta$ ``G-band''
to derive C abundances via spectrum syntheses.
Significant G-band absorption occurs in the 4200$-$4400~\AA\ spectral
range, and we estimated C abundances from three regions 
(4300$-$4308~\AA, 4308$-$4315~\AA, and 4322$-$4327~\AA) and averaged
the results.
The 4308$-$4315~\AA\ region proved to be the most reliable C abundance
indicator, due to greater atomic-line contamination in the other
two spectral regions.
The synthetic spectrum line list was formed with \iso{12}{CH} and 
\iso{13}{CH} lines from Plez (private communication; see, e.g., 
\citealt{hil02}, \citealt{ple05}) and atomic lines from the
\cite{kur11} compendium.
The CH lines are plentiful and relatively strong throughout the
atmospheric parameter domain of our program stars.
As an example of the CH observed/synthetic spectrum matches, we show
in Figure~\ref{HIP54048_CH} the 4308$-$4315~\AA\ spectral region
in a typical program star.
Generally more reliable C abundances were obtained from the CH band
syntheses than those from the \ion{C}{1} $EW$ measurements, especially
among the cooler (\teff~$<$~5500~K) stars.

\begin{figure}
\epsscale{1.00}
\plotone{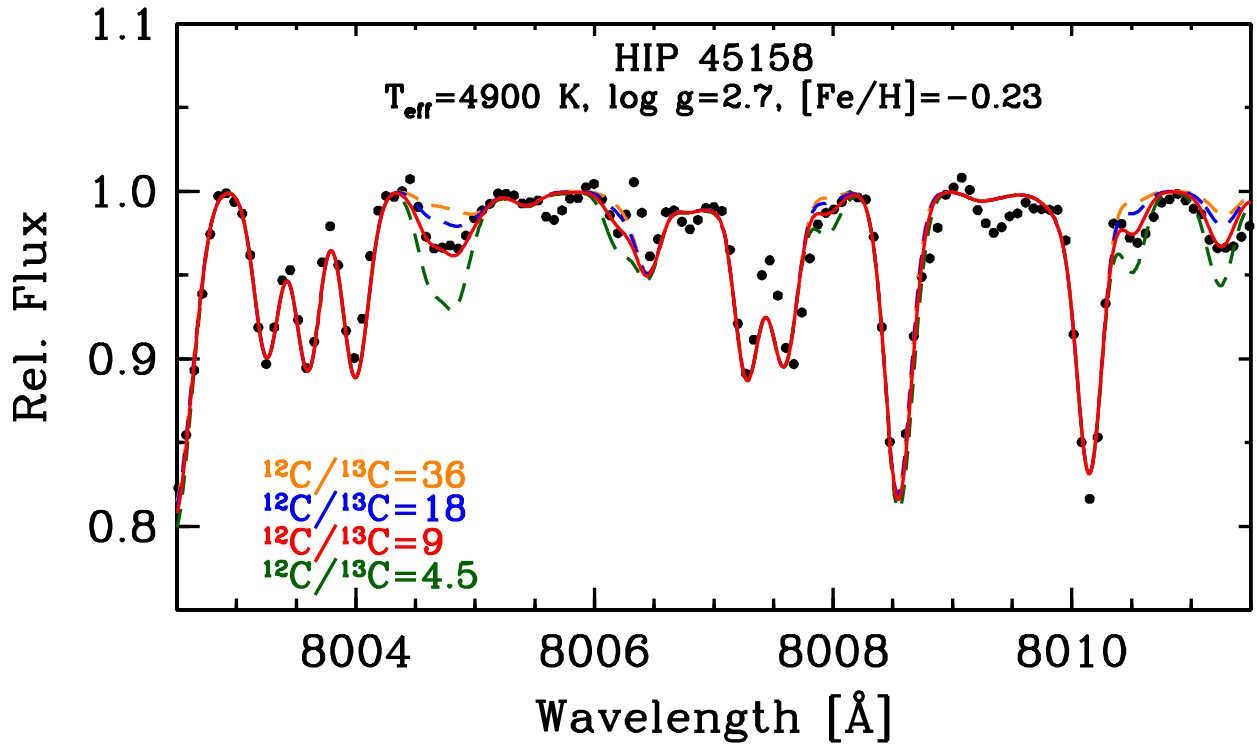}
\caption{\footnotesize
         Comparison of synthetic and observed spectrum of HIP 45158 for 
         observed $^{13}$CN feature around 8004 \AA. 
         Best fit is illustrated by a red solid line for \ciso~=~9. 
         Other trials are also shown for \ciso~values of 36, 18 and 4.5
         (dashed lines).
\label{HIP45158_CN} 
}
\end{figure}

\textbf{{CN:}} We used synthetic spectrum calculations of 
{\iso{12}CN} and \iso{13}{CN} $A^2\Pi-X^2\Sigma$ red system lines in 
the 7995$-$8040~\AA\ region to determine N abundances and \ciso\ ratios. 
The synthesis line lists were taken from \cite{kur11}.
A triplet of \iso{13}{CN} lines near 8004.7~\AA\ was the primary 
\ciso\ indicator, but several other, usually weaker features 
(e.g., at 8007.9, 8010.4 and 8011.2~\AA) were used for confirmation.
In Figure~\ref{HIP45158_CN}, we give an example of a synthetic spectrum 
fit to the  $^{12}$CN and $^{13}$CN features in the 8002.5$-$8011.5~\AA\
region.
Derived \ciso~ratios for our program stars are given in Table~\ref{tab-CN}.
C-N-O abundances are bound to each other through molecular
equilibrium, which becomes more important towards lower temperatures.
The accuracy of N abundances especially depend on the accuracy of C 
abundances through CN formation.
Note that our line list produced a systematic offset of $\sim$0.15
in the solar N abundance (Table~\ref{tab-sun}) compared to 
\cite{asp09}.
This would suggest that perhaps we have an offset of about $+$0.15~dex 
in our N abundances.

\textbf{[\ion{O}{1}]}: There are two ground-state forbidden transitions
upon which are based most of the O abundances in red giant stars.
We analyzed only the 6300.3~\AA; its companion at 6363.8~\AA\ is much
weaker and suffers from large amounts of CN contamination.
The 6300~\AA\ line has a very accurate transition probability
(\citealt{all01} and references therein).
The 6300~\AA\ [\ion{O}{1}] line is blended with a \ion{Ni}{1} line 
at 6300.34~\AA; see \citeauthor{all01}
One needs to remove its contribution carefully during the modeling of 
the [\ion{O}{1}] spectral region around 6300~\AA. 
Of some importance is the \ion{Ni}{1} oscillator strength; we use
the value given by \cite{johan03}.

\textbf{\ion{O}{1}}: In the spectra of our warmer stars the forbidden
line is too weak to yield reliable abundances, so we also used the 
often-analyzed very high-excitation 7770~\AA\ triplet. 
For these lines we adopted $gf$-values from the NIST database 
(\citealt{ral11}).
Unfortunately, it is well known that the \ion{O}{1} triplet lines are
subject to NLTE effects, in the sense that their LTE-based abundances
are always too large ($\lesssim$ 0.2 dex, \citealt{grat99}, 
\citealt{bensby04}) compared to abundances from the forbidden lines. 
In order to correct for the NLTE effects, we followed a 
similar approach to the NLTE-correction method of \cite{bensby04}. 
That is, we applied a ``robust regression" analysis and derived
the following equation for [O$_{6300}$/O$_{7774}$]$_{\rm cor}$ 
as a function of \teff, \logg~and [Fe/H]:

\begin{math}
{\rm [O_{6300}/O_{7774}]_{cor}}=-0.293(\pm0.027)+0.429(\pm0.839).{\rm log}(\frac{\teff_{\star}}{\teff_{\sun}})\\
~~~~~~~~~~~~~~~~~~~~~~~~~~~~~~~ -0.232(\pm0.033).{\rm log}(\frac{\logg_{\star}}{\logg_{\sun}})\\
~~~~~~~~~~~~~~~~~~~~~~~~~~~~~~~ +0.136(\pm0.055).{\rm [Fe/H]}
\end{math}

\begin{figure}
\epsscale{0.90}
\plotone{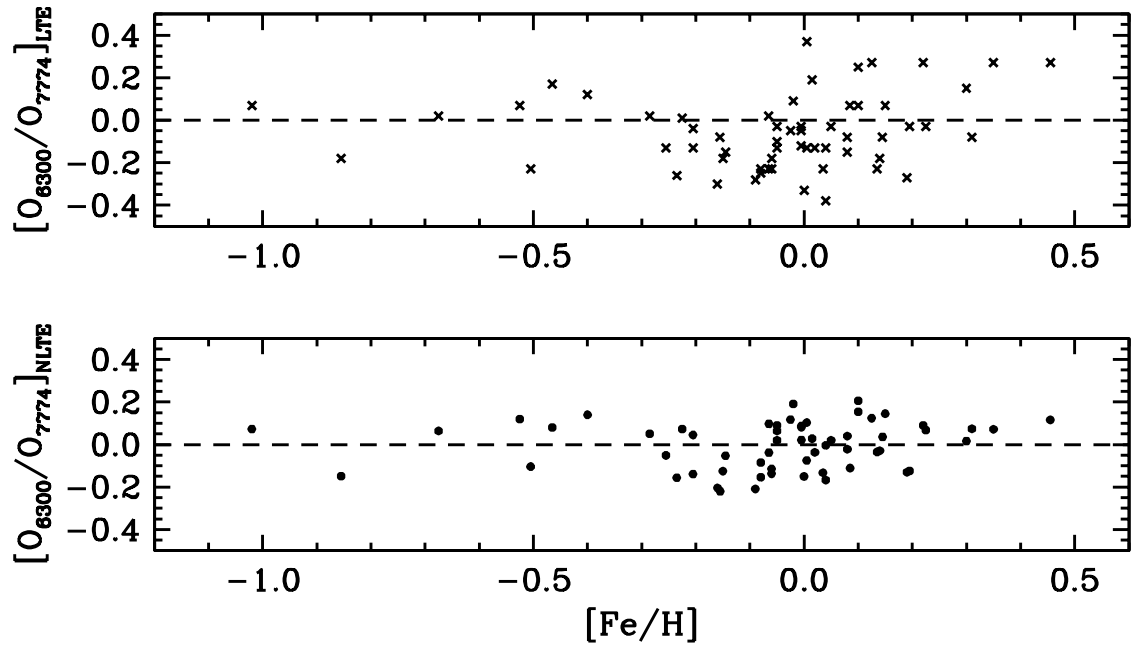}
\caption{\footnotesize
         Application of NLTE correction to the \ion{O}{1} triplet. Upper panel (crosses) shows before, bottom panel (dots) shows after the correction applied. 
\label{del_OFe} 
}
\end{figure}

Here, [O$_{6300}$/O$_{7774}$]$_{\rm cor}$ represents the difference 
between the oxygen abundances gathered from the [\ion{O}{1}]$_{6300}$ line
and the [\ion{O}{1}]$_{7774}$ triplet lines.
Then we simply subtracted this difference from [O/H]$_{7774}$ abundances 
in order to obtain NLTE-corrected [O/H]$^{\rm NLTE}_{7774}$ values. 
In Figure~\ref{del_OFe}, we compare the deviation of the
differences between [O$_{6300}$/O$_{7774}$]$_{\rm LTE}$ and 
[O$_{6300}$/O$_{7774}$]$_{\rm NLTE}$.
The mean and the standard deviation of the correlation after correction
are 0.05~dex and $\sigma$~$\simeq$~$\pm$0.12~dex, respectively.

\begin{figure}
\epsscale{1.00}
\plotone{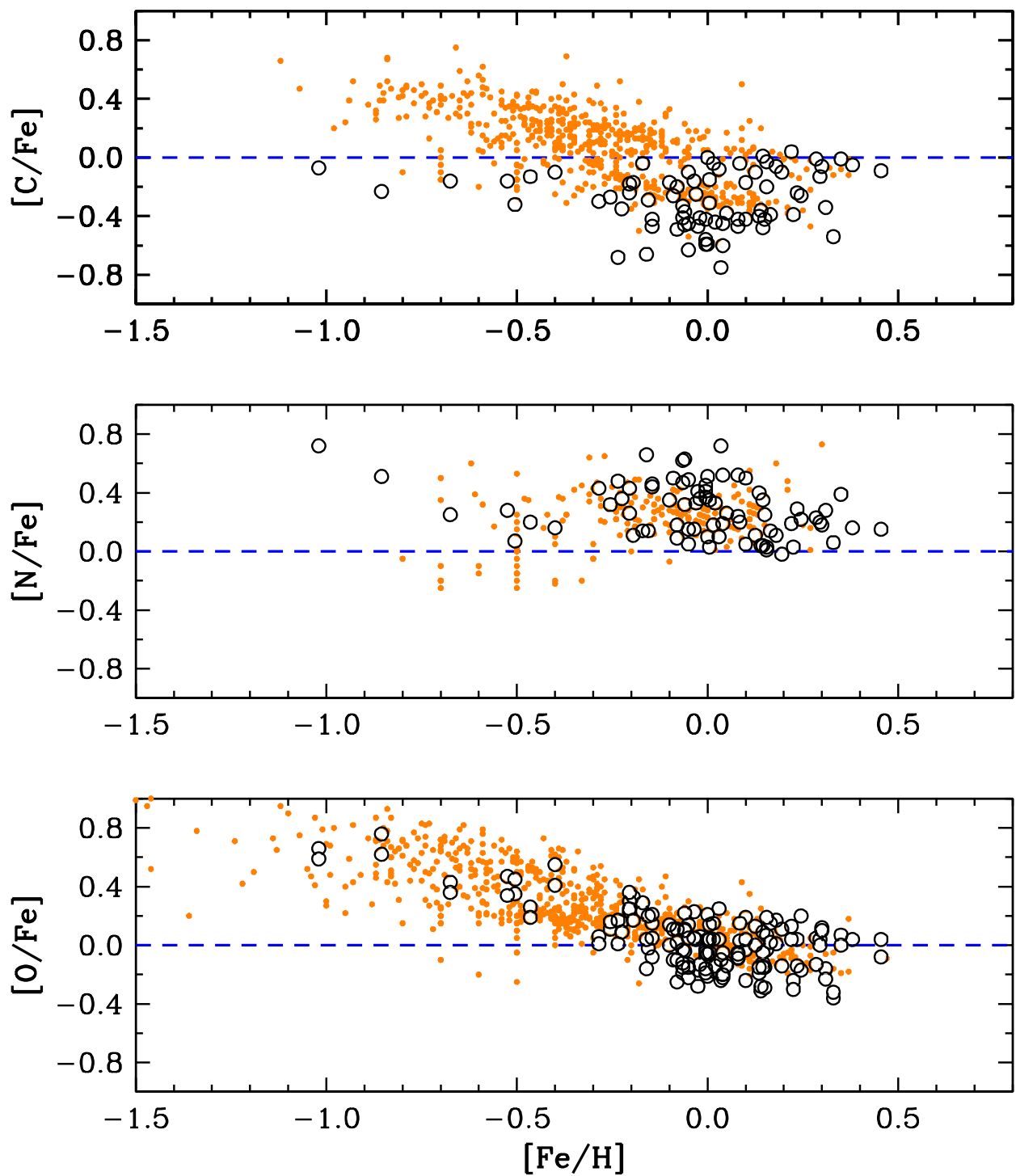}
\caption{\footnotesize
         Relative C, N and O abundances in our program stars
         (black open circles) and in stars in the literature
         (orange dots).
         All results are based on lines of \ion{C}{1}$_{CH}$, 
         \ion{N}{1} from CN8020~\AA region and \ion{O}{1}$_{6300}$ as well 
         as NLTE-corrected \ion{O}{1} from 7774 triplets species.
         The data from the literature are adopted without adjustment
         from  \cite{lamb81}, \cite{cott86}, \cite{tau01}, \cite{bensby03}, \cite{reddy03,reddy06} 
and \cite{mishenina06}.
\label{CNOC} 
}
\end{figure}

\begin{figure}
\epsscale{1.00}
\plotone{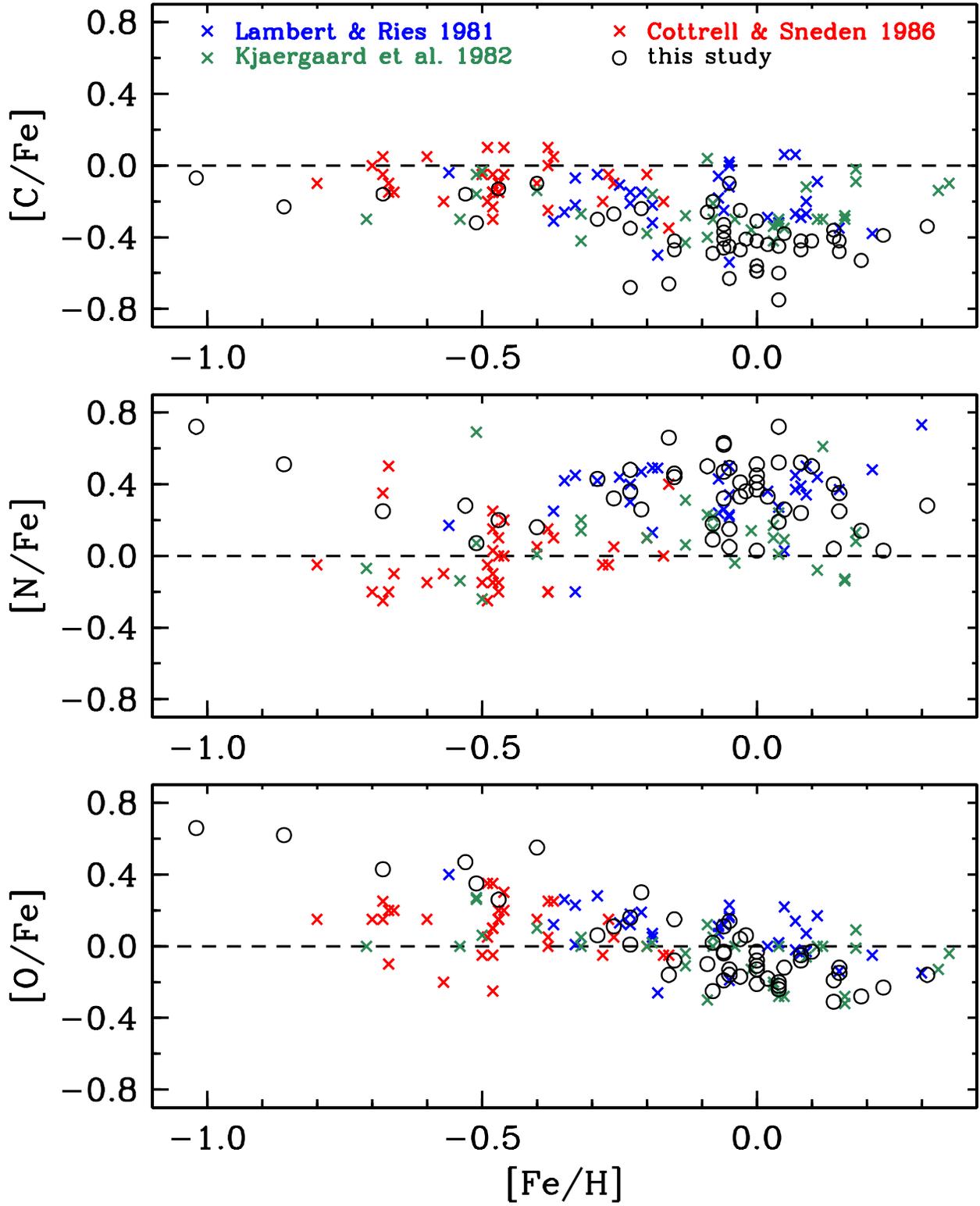}
\caption{\footnotesize
         The C, N and O abundances in our program stars (open circles) 
         and in samples of evolved (red giant) stars in the literature 
         (crosses).
         Different colors have been used to denote results from other
         studies, according to the legend given in the top panel.
\label{cnoevolved} 
}
\end{figure}

In Figure~\ref{CNOC}, we plot our CNO abundances along with 
previously reported values, using the same comparison samples. 
While the derived mean O and N abundances of our stars are
are in good agreement with literature mean values at similar metallicities,
our C abundances are somewhat lower those from previous studies.
But the comparison samples in this case are not the best, because
the literature data are dominated by main sequence and subgiant stars, 
while our program stars are clearly much more evolved.
Therefore in Figure~\ref{cnoevolved} we repeat the exercise
of the previous figure, but with two differences.
First, we have eliminated the stars of our sample that are probably
not chemically mixed (those with T~$\gtrsim$~5400~K
and/or \logg~$\gtrsim$~3.5).
Second, we added literature CNO abundance results only from studies that
concentrated on red giant stars.
Our C abundances are more in accord with those in other samples
of evolved giants.

Thick disk giants ([Fe/H]$<$--0.25) in general show higher 
C and O abundances than their thin disk counterparts. 
The O enrichment is indicative of larger Type~II SNe contributions
to thick disk than thin disk stars, combined with no depletion in ON-cycle 
H-fusion (interior temperatures are too low) in the low-mass thick-disk 
giants that we observe.
The higher C in thick disk stars suggests that the lower-metallicity,
relatively low-mass thick-disk giants have relatively shallow convective 
envelopes, and dredge-up has failed to reach the interior fusion zones 
where the CN-cycle has run to completion.
This echos the discussion in \citealt{cott86}, who concluded that in their 
old disk giants, ``only the coolest (outer) portions of the CNO-processed 
hydrogen burning shell, where \iso{12}{C}($p,\gamma$)\iso{13}{C} has 
taken place, have been convectively mixed into outer layers''.
Theoretical support for this notion was given in \cite{sneden86}, whose
Table~6 gives D. A. VandenBerg's predicted C/N ratios from standard 
evolutionary model stars that have experienced first dredge-up. 
It is clear that as as mass and metallicity decrease, model stars
exhibit decreasing surface C and N abundance changes.

\textbf{\ion{Li}{1}:} Unless the Li abundance is extraordinarily large,
only the 6707.8~\AA\ resonance line of the neutral species can be 
detected in cool stars.
This ground-state doublet transition has a well-determined $gf$ value;
see the summary of experimental and theoretical studies in \cite{smi98}.
The transition is complex, with both isotopic and hyperfine 
substructure. 
We adopted the parameters given in the line compendium of 
\cite{kur11}\footnote{http://kurucz.harvard.edu/linelists.html}.

We were able to measure {\ion{Li}{1}} abundances for some 
of our targets (see Table~\ref{tab-Li}). 
Since Li line strength is a severe function of \teff, for stars with
undetectable Li here are approximate abundance upper limits as 
guidelines for interpretation: 
for stars with \logg\ $\geq$ 3.5 and T $\geq$ 5500, log$\epsilon$(Li)~$<$~1,
and for those with \logg\ $<$ 3.0 and T $\leq$ 5100, 
log$\epsilon$(Li)~$<$~+0.5 (with the limit of course  decreasing 
towards lower temperatures). 
The Li abundances decrease as \teff\ and  \logg\ decrease.
This is as expected in normal stellar evolution: during the path from
main sequence to red giant phases, the deepening convective envelopes 
bring Li down from the surface layers where it can be destroyed easily
at relatively low fusion-zone temperatures.
Program stars with higher surface Li abundances have undetectable \iso{13}{C},
consistent with their chemically un-evolved status.
Six stars of our sample have low \ciso\ values, but detectable 
\ion{Li}{1} lines; these all appear to be plausible RHB and RC 
candidates (see Table~\ref{tab-abun}). 
For these stars, $<$log~$\epsilon$(Li)$>$~$\sim$~0.8.
Standard evolution models (e.g., those without extra mixing mechanisms
seeming needed for Population~II giants) predict surface Li depletion
by factors up to about 30 (\citealt{iben67a}).
If a star begins main-sequence life with the present interstellar medium
Li abundance (log~$\epsilon$(Li)~$\sim$~+3.0, e.g. \citealt{grev98}) 
then through ordinary stellar evolution it should exhibit 
log~$\epsilon$(Li)~$\lesssim$~+1.5.
None of our evolved program stars have Li abundances near to this value,
indicating either that their natal Li contents were not as high as the
present ISM value, or that extra mixing during their lives destroyed
Li beyond the standard predictions.

\section{KINEMATICS\label{motions}}

Among the overall sample of 76 stars, we have distance 
information for 58 of them to facilitate a kinematical investigation.
In order to calculate stellar space velocities, we have collected 
parallaxes and proper motions, as well as their errors, from \cite{lee07}, 
who recently published new reduction results of HIPPARCOS data. 
We either adopted the RVs already available in the literature or we 
measured new values as described in \S\ref{data}.
We used the matrix equations given by \cite{john87} and calculated 
the Galactic velocity components $U_{\rm{LSR}}$ (positive toward the 
Galactic center, $V_{\rm{LSR}}$ (positive in the direction of the 
Galactic rotation), $W_{\rm{LSR}}$ (positive toward the NGP), 
and their uncertainties
($\sigma_{U_{\rm{LSR}}}$, $\sigma_{V_{\rm{LSR}}}$,  $\sigma_{W_{\rm{LSR}}}$) 
with respect to the Local Standard of Rest (LSR). 
Correction for the solar motion was made by adopting the values 
$(U, V, W)_{\sun}$ = (+10.00, +5.25, +7.17) from \cite{deh98}. 
The results are given in Table~\ref{tab-kin}.

\begin{figure}
\epsscale{1.00}
\plotone{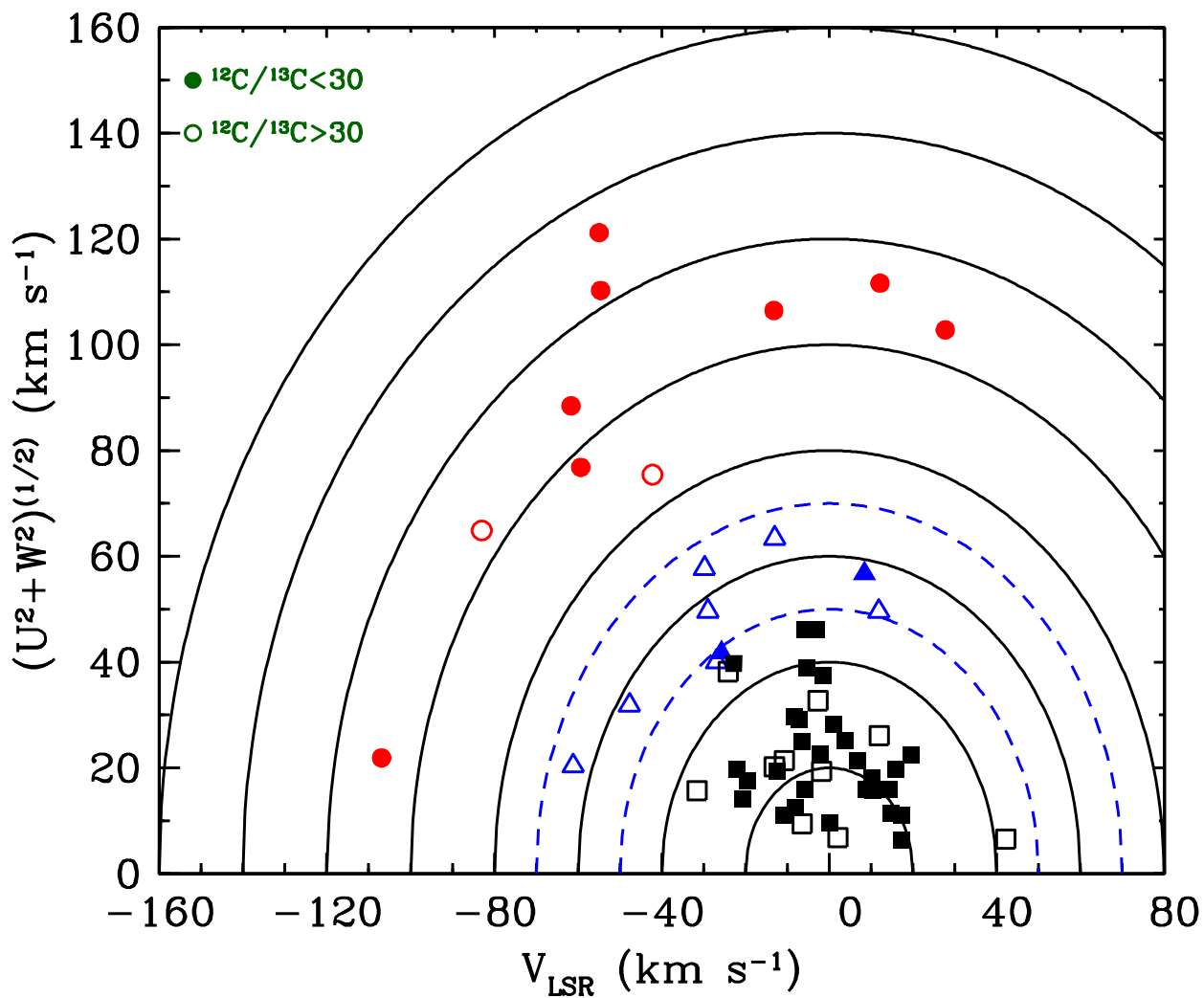}
\caption{\footnotesize
         Toomre diagram of the program stars. 
         Solid lines indicate $V_{\rm{tot}}$ in steps of 20 \kmsec.
         The dashed lines at 50 and 70 \kmsec\, denotes the approximate
         thin-thick disk separation. Thin, thin/thick and thick disk stars are represented by 
        (black) squares, (blue) triangles and (red) circles.
         Filled and open symbols represent whether \ciso\ is 
         detected in the spectrum of the related star (representative filled and open symbols 
         and their meaning are also shown in the figure).  
\label{spvel}
}
\end{figure}

In Figure~\ref{spvel} we show these kinematics in a Toomre diagram.
This kind of plot gives the kinematical distribution of the stars in 
terms of their combined vertical and radial kinetic energies as 
a function of their rotational energies.
The solid curves of Figure~\ref{spvel} are lines of constant 
total space velocity, 
$V_{\rm{tot}}=({{U}^2}_{\rm{LSR}}+{{V}^2}_{\rm{LSR}}+{{W}^2}_{\rm{LSR}})^{1/2}$.
The dashed line represents the $V_{\rm{tot}}$ = 70 \kmsec, which 
approximately separates the thin and thick disk populations of the 
Galaxy (e.g. \citealt{bensby09}, \citealt{nissen04}). 
Stars with total space velocities of 
70~$<$ $V_{\rm{tot}}$~$<$ 180~\kmsec are 
considered to be probable thick disk, those in the range 
50~$<$ $V_{\rm{tot}}$~$<$ 70 \kmsec are called transition (thin/thick)
objects, and those with $V_{\rm{tot}}$~$<$ 50~\kmsec are 
probable thin disk members of the Galaxy.

Inspection of Figure~\ref{spvel} reveals that our 
sample is dominated by thin disk stars.
Out of 58 stars with kinematic information, 39 (the squares) 
probably reside in the thin disk, 10 probably are in the thick disk,
and 9 are kinematic transition objects whose thin/thick status is
ambiguous.
In Figure~\ref{spvel} we also use filled and open version of the symbols 
to indicate the detection status of $^{13}$CN features in individual 
stars: those with \ciso~$<$~30, and those with no detection of 
$^{13}$CN (thus \ciso~$>$~30), respectively.
We will consider the carbon isotopic ratio distribution further in 
\S\ref{interp}.

\begin{figure}
\epsscale{1.00}
\plotone{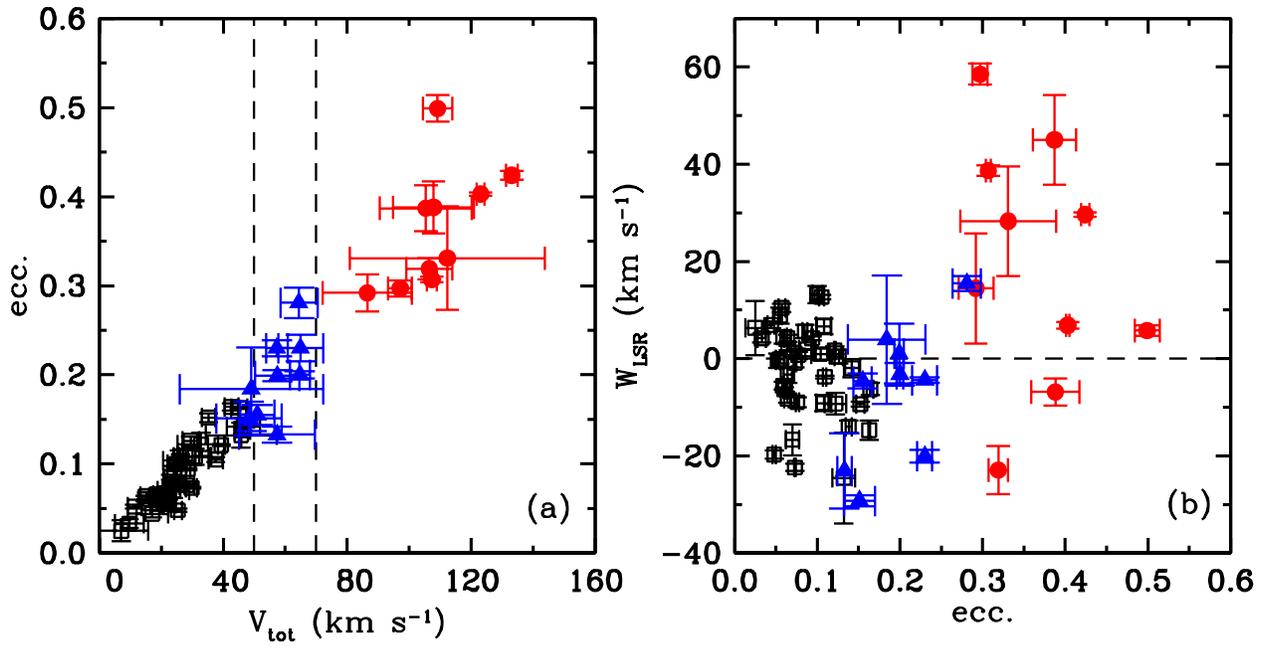}
\caption{\footnotesize
         Panel $\it (a)$: correlation of Galactic orbital eccentricity 
         $ecc.$ with total space velocity $V_{\rm{tot}}$.
         Panel $\it (b)$: correlation of vertical velocity
         component $W_{\rm{LSR}}$ with eccentricity.
         The symbol colors are the same as those of Figure~\ref{spvel}.
\label{veleccC}
}
\end{figure}

Looking at the kinematics of our stars in more detail, in 
Figure~\ref{veleccC} we show the correlations between total space 
velocity $V_{\rm{tot}}$ and eccentricity, and between the
velocity component of $W_{\rm{LSR}}$ and eccentricity.
The eccentricities were calculated using the method described in \cite{dana99}.
The solar circle radius of R$_{0}$~= 8.0~kpc around the Galactic center 
and an LSR rotation velocity of $\Theta$~= 220~\kmsec\ were adopted.
As expected, the total velocity is strongly correlated with
eccentricity in our sample.
However, the relationship between the vertical velocity and
eccentricity is not obvious beyond increased scatter in $V_{\rm{tot}}$
at large eccentricities.
Several of our thick disk stars have low $\vert{W_{\rm{LSR}}}\vert$ 
velocities ($\lesssim$20 \kmsec).
Given that $W_{\rm{LSR}}$ is proportional to z$_{max}$, the maximum 
vertical distance of a star to the Galactic plane, these particular stars 
with $W_{\rm{LSR}}$ apparently reside close to the Galactic plane.

\begin{figure}
\epsscale{1.00}
\plotone{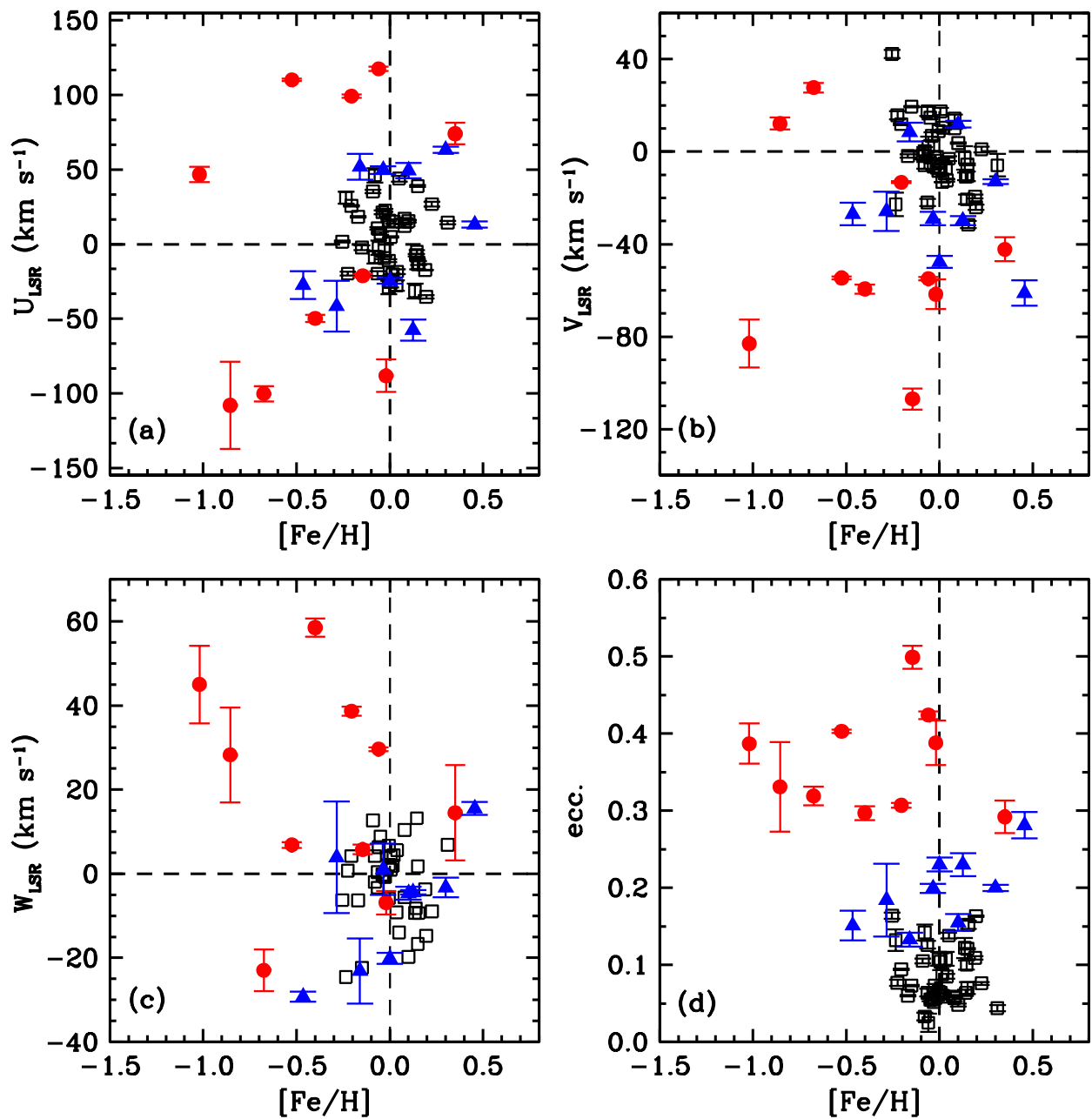}
\caption{\footnotesize
         Correlations of kinematic components with [Fe/H] metallicity:
         $U_{\rm{LSR}}$ in panel $\it(a)$; $V_{\rm{LSR}}$ in panel $\it(b)$; 
         $W_{\rm{LSR}}$ in panel $\it(c)$; and eccentricity in panel $\it(d)$.
         The symbol colors are the same as those of Figure~\ref{spvel}.
\label{FeHkinC}
}
\end{figure}

The space motions and derived metallicities of our program stars are related.
In Figure~\ref{FeHkinC}, we plot the three space velocity 
components against [Fe/H] in panels {\it (a)}, {\it (b)}, 
and {\it (c)}.
The larger dispersions of the $U_{\rm{LSR}}$, $V_{\rm{LSR}}$, 
and $W_{\rm{LSR}}$ velocities in the true thick-disk stars is 
clear from inspection of these panels.
Useful velocity limits in order to separate the thin and thick 
disk stars appear to be $\vert{W_{\rm{LSR}}}\vert$~$\approx$ $+$20~\kmsec\ 
and $V{_{\rm{LSR}}}$~$<$ $-$40~\kmsec. 
However, all three velocity components that comprise total space velocity 
must be taken into account in assessing the Galactic membership of 
individual stars. 
In panel {\it (d)} of Figure~\ref{FeHkinC} we plot [Fe/H]
versus eccentricity, and it is clear that lower metallicity
stars in our sample generally have more eccentric orbits. 
This is similar to the relation recently suggested by \cite{lee11}.
However, their study considered only thick-disk stars with subsolar
metallicities ([Fe/H]~$<$~$-$0.3).
We stress here that stars with thick-disk kinematics (including
eccentricity) but [Fe/H]~$>$~$-$0.3 have no obvious metallicity-kinematics
connections.

\begin{figure}
\epsscale{1.00}
\plotone{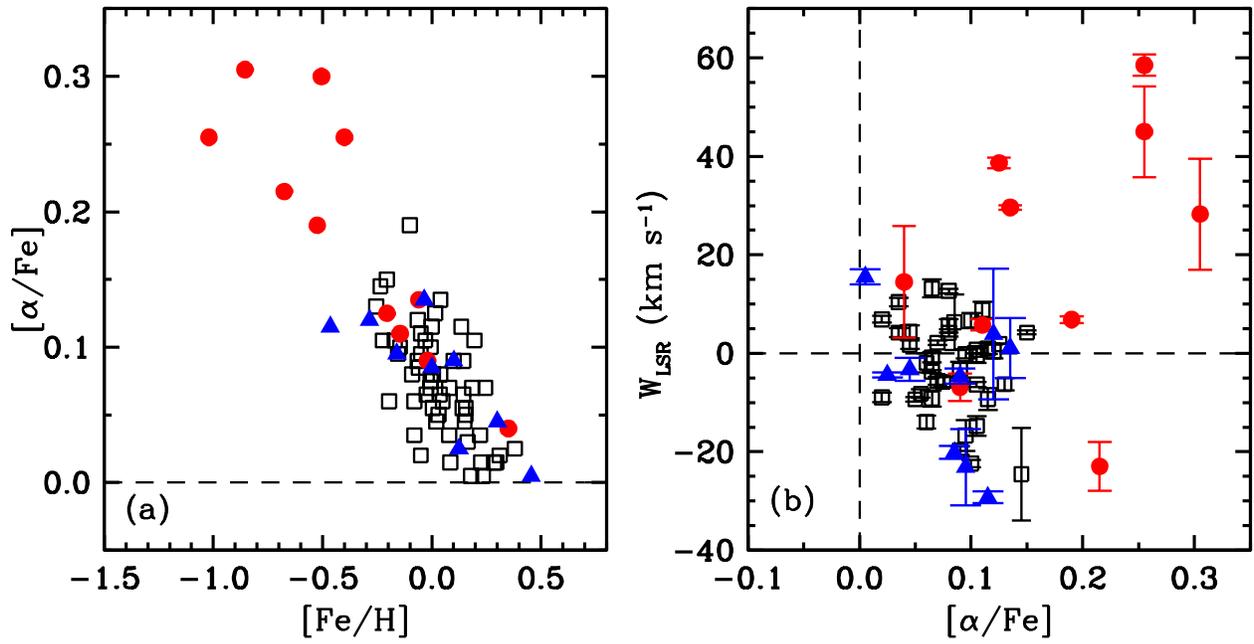}
\caption{\footnotesize
         Panel $\it (a)$: correlation of relative [$\alpha$/Fe]
         abundance ratios with respect to metallicity [Fe/H].
         Panel $\it (b)$: comparison of vertical velocity component
         $W_{\rm{LSR}}$ with [$\alpha$/Fe] values.
         The symbol colors are the same as those of Figure~\ref{spvel}.
\label{alphavelC}
}
\end{figure}

In Figure~\ref{alphavelC} we explore the relationship between our
$\alpha$-element abundances ([$\alpha$/Fe] = ([Ca/Fe]+[Si/Fe])/2), 
metallicity, and kinematics.
In panel {\it (a)} we correlate [Fe/H] with [$\alpha$/Fe].
The well-documented rise in relative $\alpha$-element abundances
with decreasing metallicity in disk populations is reproduced in our
relatively small sample. 
All of our thick-disk stars with [Fe/H]~$<$~$-$0.3 have 
[$\alpha$/Fe]~$\gtrsim$~$+$0.2.
Although our sample is small, there seems to be a clear separation 
between the thick and thin disk stars for the metallicities of 
[Fe/H] $<$ --0.3 as indicated in larger surveys (e.g. \citealt{reddy06}).
Stars with [Fe/H] $>$ --0.3 show similar [$\alpha$/Fe] ratios regardless of 
their membership either in the thick or the thin disk.
In panel {\it (b)} we plot the [$\alpha$/Fe] versus metallicity.
This relationship can be compared with the $W_{\rm{LSR}}$ versus
[Fe/H] plot shown in panel {\it (d)} of Figure~\ref{FeHkinC}.
They tell essentially the same story, linking in our sample the
kinematics, metallicities, and $\alpha$-element abundances of our sample.

\section{EVOLUTIONARY STATUS\label{evstat}}

\begin{figure}
\epsscale{1.00}
\plotone{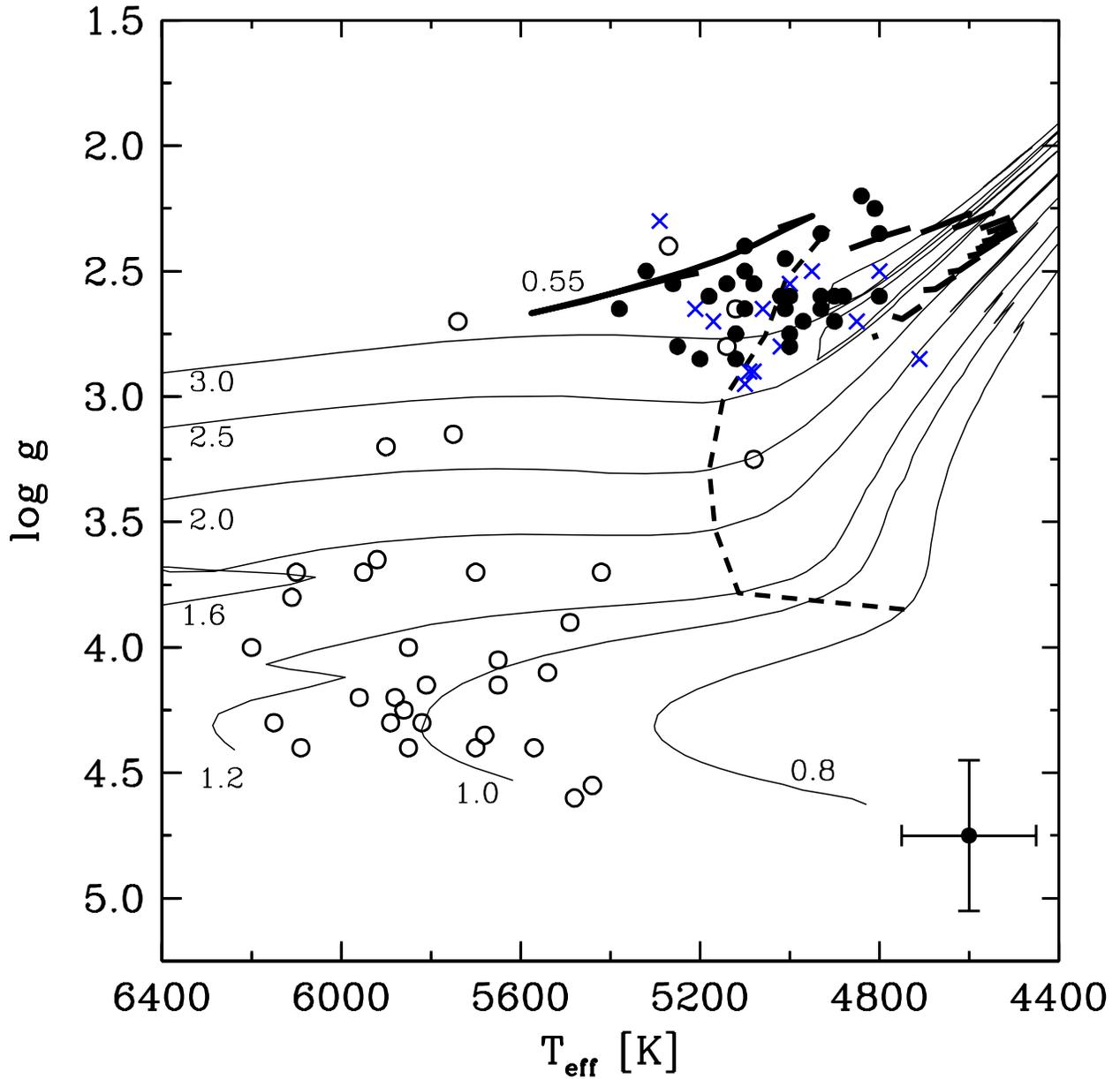}
\caption{\footnotesize
         Evolutionary states of our program stars are shown on the 
         spectroscopic \logg-\teff~diagram. Stars with \ciso\,$\leq$ 20, 
         20 $<$\ciso\,$\leq$ 30 and \ciso\,$>$ 30 (no detection) are shown by filled circles,
         (blue) crosses and open circles, respectively. 
\label{HR}
}
\end{figure}

We estimated the probable evolutionary stages of our program 
stars by taking into account their loci in the \teff--\logg\ plane,
their absolute magnitudes (luminosities), and their \ciso\ values 
derived from our spectral analysis.
In Figure~\ref{HR} we show the \teff$-$\logg\ diagram, using different
symbols to denote stars with and without detected \iso{13}{CN} spectral
features.
Although we selected our targets as luminosity class$-$III stars,
our atmospheric analyses revealed about 1/3 of our sample to be either 
MS or SGB stars
These are evident in the figure as higher temperature, higher gravity
stars with weak/absent \iso{13}{CN}.

In Figure~\ref{HR} we have also shown theoretical evolutionary tracks 
from \cite{bert08} (Y=0.26, Z=0.02; note that we do have a few low 
metallicity stars in our sample but for simplicity we plot only tracks 
with a single metallicity).
The evolutionary-track masses displayed here are from 0.8 to 3 M$_{\sun}$. 
For tracks with 0.8$-$2 M$_{\sun}$ the evolutionary phases go up 
to RGB tip (at lower temperatures than displayed here), 
while the 2.5 and 3 M$_{\sun}$ tracks proceed up to the 
first thermal pulse stage.
The base of the RGB, including masses up to 4 M$_{\sun}$, is also 
shown with a dashed line, which is constructed from the data
of \cite{bert08}.
We have also included HB evolutionary tracks for the mass range 
0.55$-$2 M$_{\sun}$ (denoted by the thick black lines). 
Only 0.55 M$_{\sun}$ HB track evolves up to the early AGB phase. 
Other HB tracks start from the zero-age horizontal branch (ZAHB) and 
go up to the maximum He-burning phase. 
HB tracks with masses higher than 0.55 M$_{\sun}$ make an $arc$ shape 
in Figure~\ref{HR}.
This arc covers a temperature range 
4500~K~$\lesssim$ \teff~$\lesssim$ 4900~K, and thus essentially
defines the RC region. 
RC stars are the reddest HB stars with higher masses than normal
RHB stars. 
Both of them are at the core He-burning stage and all belong to the 
RHB class in general.

In \S\ref{errors} we compared spectroscopic surface gravities with 
theoretical ones (Figure~\ref{logg}), suggesting that, for most of 
our targets, the masses are $\sim$2M$_{\sun}$. 
This is consistent with the inferred evolutionary-track masses of 
the warmer (\teff~$>$~5300~K), higher gravity (\logg~$>$~3.0) stars 
in Figure~\ref{HR}.
The SG stars should not yet have convectively-mixed envelopes,
so it is not surprising that these objects also do not have 
detectable $^{13}$CN lines and hence \ciso~$>$~30 (e.g. \citealt{thoren04}).

For almost all of the stars with \teff~$<$~5300~K, the 
masses inferred from the evolutionary tracks in Figure~\ref{HR} 
are $\gtrsim$ 3 M$_{\sun}$.
These implied masses are much larger than those of the warmer 
MS/SG program stars, and also are inconsistent with our earlier 
(\S\ref{errors}) assertion that almost all of our stars have actual 
masses $\lesssim$ 2 M$_{\sun}$.
The implication is that our lower temperature stars are not SG stars 
on their first ascent up the RGB, but instead are post-RGB stars
in the helium-burning RHB evolutionary stage.
It is possible for RHBs to be confused with high-mass SGs, which can
have similar absolute magnitudes to the RHBs. 
But such SG stars should be rare because the timescales for 
passage of $\gtrsim$ 3 M$_{\sun}$ stars through the 4800$-$5300~K 
temperature domain ($\sim$10$^{5}$--10$^{6}$ years) are short compared 
to the RHB He-burning timescales ($\sim$10$^{8}$ years) of less
massive (thus probably more plentiful) stars.
Our cooler program stars in general ought to be true members of the 
field RHB population.

For each program star we have estimate a probable evolutionary stage,
taking into account its position in the \teff-\logg\ plane from arguments 
presented above, its absolute magnitude when parallax data are available,
and its \ciso\ ratio.
These estimates are given in the last column of the Table~\ref{tab-abun}.
We suggest that there are 18 RHB stars in our sample
(5 thick disk and 13 thin disk), along with several RC and RC/RHB stars, 
most of them members of the thin disk.

\section{SUMMARY AND DISCUSSION\label{interp}}

In this study we have determined atmospheric parameters and
chemical compositions of field RHB candidates selected simply from their
colors and absolute magnitudes.
The original goal was to increase the sample size of bona-fide RHB stars, 
and to try to understand the physical processes involved in the existence 
of these relatively rare highly-evolved objects. 
We deliberately avoided kinematic biases by employing only photometric and 
spectral type information in selecting stars for observation.

We gathered high resolution, high S/N spectra of 129 candidate field RHB stars. 
Preliminary analyses eliminated 53 candidates from further consideration 
because they proved to be anomalously broad-lined rapidly rotating stars, 
or double-lined spectroscopic binaries, or stars for which we could
not deduce reliable atmospheric parameters.
For the remaining 76 program stars, we first determined values
of \teff, \logg, \vmicro, and [Fe/H].
Then we derived abundances for $\alpha$ elements Si and Ca, 
neutron-capture elements La and Eu, proton capture elements Li, C, N, and O.
We also determined \ciso\ isotopic ratios, 
because detection of \iso{13}{C} provides strong evidence of
CN-cycle H-fusion and mixing associated with evolved stars.
The derived fundamental stellar parameters \teff\ and \logg, along with
the \ciso\ isotopic ratios, were used to estimate the evolutionary stages 
of our program stars. 
This evaluation suggests that about 20\% of 76 program stars are 
true RHB members.

We then computed space velocity components for all of the program stars
with available distances, proper motions, and radial velocities (either
from the literature or from our own spectra).
For each star we estimated Galactic population membership (thin disk, 
thick disk, and thin/thick transition) only from the kinematics.
We examined correlations between kinematics and (a) overall metallicity,
and (b) relative abundance ratios of the $\alpha$ elements (confirming
these with abundance ratios of the neutron-capture elements).
Even with our relatively small sample we recovered well-known differences
between thin and thick disk stars. 
Our thick disk stars have higher space velocities and orbital eccentricities,
lower metallicities, and larger $\alpha$-element ratios than do their
thin disk counterparts.

We did not take into account the kinematical constraints 
during the sample selection, and our moderately small survey of 76 RHB
candidates turns out to contain only five probable thick disk stars. 
These five true thick disk RHBs are very similar to others already identified
in the literature (e.g. \citealt{cott86}, \citealt{tau01}): they are 
mildly metal-poor, $\alpha$-enhanced (including high oxygen) stars that 
have low \ciso\ ratios.
They also show little evidence for carbon depletion, as discussed 
in \S\ref{pcapture}.
Such stars have relatively low masses, and their appearance on the RHB
(instead of the RC) is not surprising: mildly metal-poor globular
clusters have well-populated RHBs.

A perhaps more interesting result is that our identified RHB 
sample is dominated by thin disk, high metallicity stars.
These stars, with \teff~$>$ 5000~K, are too hot to be RC stars, which have 
temperatures lower than 4900~K. 
They also have evolved-star gravities: \logg~=~2.2$-$2.8, substantially
smaller than the gravities of subgiants and main-sequence stars
(Figure~\ref{HR}).
They show no obvious $\alpha$ and neutron-capture abundance anomalies 
compared to other thin-disk samples.
Their proton-capture abundances are unremarkable compared to normal
thin-disk RGB stars.
In particular, they exhibit [O/Fe]~$\sim$~0 and [C/Fe]~$\sim$~$-$0.4,
and often low \ciso\ values, all consistent with expectations from
past studies.
But their residence in the RHB cannot be easily understood through standard
stellar evolution considerations.

Both thick and the thin disk RHBs have a range of \ciso\ ratios
ranging from 5 to 30 (Table~\ref{tab-abun}). 
Before discussing the \ciso\ ratios of our stars in detail, we review 
the physical processes that can affect \ciso\ ratios during 
stellar evolution in the next paragraphs.

As a low mass star ($<$ 2.25 M$_{\sun}$) evolves past the MS and SG
evolutionary stages, the first dredge-up starts at the base of the 
RGB (e.g. \citealt{iben64,iben65,iben67b}), accompanied by convective 
envelope expansion towards inner layers of the star. 
This will bring CN-cycle processed material up to the outer layers, 
thus, resulting in surface abundance alterations of the LiCNO group.
Fragile Li is severely depleted first.
As the convection extends into deeper inner regions, it passes 
through the transition region which separates the region of $^{12}$C 
converted into $^{14}$N.
In standard dredge-up theories, mixing of the processed and unprocessed 
elements results in depletion of surface $^{12}$C, lowering the \ciso~ratio 
to $\sim$20$-$30 from its (assumed) initial solar value of 
$\sim$90 (\citealt{char94}, \citealt{char98}, \citealt{grat00}), while 
the surface $^{14}$N values increase. 
First dredge$-$up phenomenon leaves a mean molecular gradient 
($\mu-$barrier) behind which prevents further mixing (e.g, \citealt{char98}).

However, previous observational studies have shown that 
\ciso~ratios as low as the CN$-$cycle equilibrium value of 3.4 
(e.g. \citealt{sneden86}, \citealt{cott86}, \citealt{grat00}, 
\citealt{tau01}), which obviously requires a non-canonical mixing process 
during the RGB phase. 
Recent stellar evolution studies have attempted to solve this extra-mixing 
problem by offering physical mechanisms such as rotation-induced mixing 
(\citealt{zahn92}), cool bottom processing (CBP, \citealt{boot99}) and 
thermohaline instability plus rotation$-$induced mixing 
(\citealt{char07}, \citealt{char10}). 
These ideas provide an extra-mixing process that sets in after the 
``$\mu-$barrier elimination'' which occurs after the convective 
envelope base recedes towards the surface.
Then the H-burning shell eliminates the composition discontinuity left 
by the convective envelope and enables extra-mixing processes to 
come into play. 
For low mass stars, this stage of evolution is also called the
``RGB luminosity function bump (LFB)" (\citealt{grat00}, \citealt{char10}) 
at which the H-burning shell burns the newly supplied fuel as it expands 
outwards. 
This stage slows down the evolution and causes a temporary luminosity 
drop along the RGB. 
For solar metallicities and initial masses of $\leq$ 2.25 M$_{\sun}$ 
LFB appears at effective temperatures lower than 4800~K 
(e.g. \citealt{char00}, \citealt{mona11}).

As is seen in Figure~\ref{HR}, the evolved stars of our sample 
(thick+thin disk) are located approximately between 4800$-$5400~K in 
\teff\ and 2.2$-$2.8 in \logg.
The stars illustrated with filled circles have low \ciso\ values 
($\leq$~20) and they clearly reside far away from LFB.
We suggest that the stars with especially low isotopic ratios, 
\ciso\ $\leq$~10, may have evolved from lower initial masses and 
undergone a major extra-mixing processes when they passes the LFB.
The amount of the extra-mixing is related to both the initial mass and
the metallicity of the star, which may explain the deviation in the 
\ciso\ values ($\leq$~20).

Stars have masses higher than 2.25 M$_{\sun}$ ignite their central 
He before the core becomes degenerate. 
The H-burning shell which surrounds the He-burning core remains 
until the second dredge-up and never reaches the region of molecular
discontinuity left by the convective envelope during the first dredge-up
phase. 
That is why, stars with high initial masses are not expected to show 
the indicators of extra-mixing, such as low \ciso~ratios 
(\citealt{char94}, \citealt{char98}, \citealt{char07}).
In our sample, we have evolved stars with 20 $<$ \ciso~$<$ 30, which are 
close to the values suggested by canonical models (e.g. \citealt{schal92}, 
\citealt{char94}). 
These stars are shown with (blue) crosses in Figure~\ref{HR}.
Assuming that these stars did have on average larger initial masses than
the stars with low isotopic ratios, then the isotopic ratio issue is
``solved'', but leaves a problem: why are these stars now in the RHB
domain instead of the RC?
Here we suggest, but cannot prove, that the RHB stars with higher
\ciso\ ratios have undergone substantial mass loss at some stage(s) of
their evolution, leaving them with smaller present envelope masses.
Such stars will appear bluer (hotter) than RC stars.

\begin{figure}
\epsscale{0.80}
\plotone{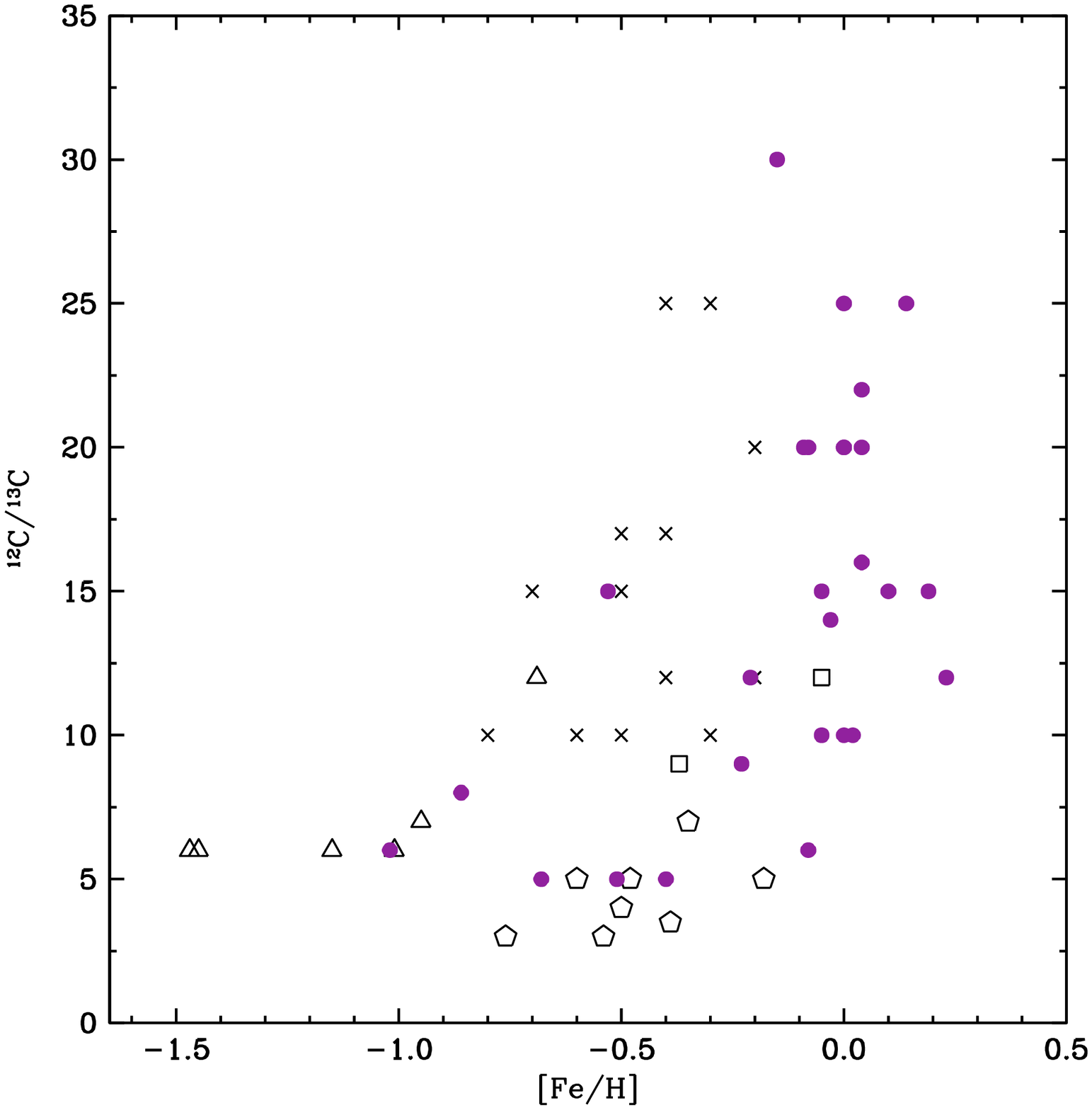}
\caption{\footnotesize
              \ciso~ratios (filled circles) as a function of the [Fe/H] 
         for RHB, RC and RC/RHB stars. 
         Other RCs and RHBs from \cite{lamb81} (open squares), 
         \cite{cott86} (crosses), \cite{grat00} (triangles) and 
         \cite{tau01} (open pentagons) are also plotted.
\label{FeH_CC}
}
\end{figure}

For further insight in Figure~\ref{FeH_CC} we correlate 
\ciso~ratios and [Fe/H] metallicities.
We include our RHB, RC, and RC/RHB stars along with the data obtained
by \cite{lamb81}, \cite{cott86}, \cite{grat00} and \cite{tau01}
for the same temperature, luminosity and surface gravity range 
(same evolutionary stages).
We have excluded SG stars and upper RGB stars in this plot.
In Figure~\ref{FeH_CC}, we see a conspicuous trend with metallicity, 
similar to the one shown in \cite{sne91}: lower metallicity RHB and RC
stars have a smaller \ciso\ range (and lower values on average) than
do similar objects of higher metallicity.
As discussed in, e.g. \cite{kal09} and \cite{cate09}, mass loss in evolved
stars increases rapidly with increasing metallicity. 
If the \ciso~ratio is also related to mass-loss along with mixing 
processes then for a given metallicity, stars with different mass-loss 
and mixing history must be responsible from the \ciso~fluctuations 
seen towards higher metallicities in Figure~\ref{FeH_CC}.

According to \cite{char94}, the peak of the $^{13}$C abundance 
in the inner region of a star is shifted outward towards high initial masses.
If the star evolve from high-masses ($>$ 2.25 M$_{\sun}$), then different 
mechanisms, such as winds on the RGB, for mass-loss should come into 
play other than He-flash.
Depending on how massive the star at the beginning, it might be possible
for a star to dredge up some or all of its region, where the $^{13}$C 
abundance peaks, during its RGB evolution.
Since the thin disk stars are more metal-rich compared to thick disk 
and halo stars, the thin disk stars with higher metallicities or higher 
masses or a combination of both might have very low \ciso~rates or not 
show any $^{13}$CN future depending on the mass fraction lost during the 
RGB evolution. 
High-mass evolving stars appear to have more options to produce a variety
of surface abundances.
Figure~\ref{FeH_CC} should become more illuminating if we increase the 
number of the evolved stars which are the members of the same population. 
If for example we can substantially augment the RHB thick and thin disk samples,
then we might have a better understanding of the mechanisms alters the CN ratios 
during their post-MS evolution.

We conclude by re-emphasizing our basic result: the RHB 
\teff\--~\logg\ domain of the Galactic field is populated by sub-solar 
metallicity thick-disk stars (as expected) but also by high metallicity 
thin disk stars.
We will be gathering more thin and thick disk data in order to substantially
increase the RHB samples of different populations to try to set additional
constraints on the mechanisms that may lead to stars ending up in this
relatively rare evolutionary state.

\acknowledgments

We thank Bertrand Plez for sharing his CH line lists with us,
and Andy McWilliam and Inese Ivans for making their model atmosphere
interpolation software available to us. 
MA also would like to thank The Scientific and Technological Research 
Council of Turkey (T\"{U}B\.{I}TAK) for supporting her
research by the ``Postdoctoral Research Scholarship" program.
This research has made use of NASA's Astrophysics Data System 
Bibliographic Services; the SIMBAD database and the VizieR service, 
both operated at CDS, Strasbourg, France; and the NASA/IPAC 
Extragalactic Database (NED) which is operated by the Jet Propulsion 
Laboratory, CalTech, under contract with NASA.
Support for this study has come from the U.S. National Science Foundation
grant AST~09-08978 and the Rex G. Baker, Jr. endowment to the University
of Texas Department of Astronomy.

\appendix
\section{Appendix}

{\textbf {HIP 46325}}: Super metal rich (SMR, [Fe/H] $>$ 0.2) star. 
[Fe/H]=0.46. Possible member of a moving group HR 1614 (\citealt{feltz00}). 
With a V$_{\rm tot}$ = 64 \kmsec, probable thick disk member with high metallicity.  
And probable planet host (\citealt{rob07}). 

{\textbf {HIP 38801}}: [Fe/H]=-0.06. High-temperature star. 
Shows relatively high turbulence. It has peculiarly high \ion{La}{2} 
and \ion{Eu}{2} abundances. Has $^{12}${C} deficiency and $^{14}${N} enhancement, 
indicates a more evolved star than a SG but no \ciso~ratio could be detected due 
to weak $^{13}$CH or $^{13}${CN} features.

\clearpage

\clearpage

\clearpage

\clearpage

\clearpage

\clearpage

\clearpage

\clearpage

\clearpage

\clearpage

\clearpage

\clearpage

\clearpage

\clearpage

\clearpage

\clearpage

\clearpage

\clearpage

\clearpage

\begin{center}
\begin{deluxetable}{lcccccrrrrr}
\tabletypesize{\footnotesize}
\tablewidth{0pt}
\tablecaption{Program Stars\label{tab-1}}
\tablecolumns{11}
\tablehead{
\colhead{Stars}           &
\colhead{$B$}              &
\colhead{$V$}           &
\colhead{$K_{s}$}     &
\colhead{$\bv$}     &
\colhead{$V - K_{s}$}     &
\colhead{$\pi$ (mas)}     &
\colhead{$\sigma_{\pi}$ (mas)}     &
\colhead{$\mu_{\alpha}$ (mas/yr)}     &
\colhead{$\mu_{\delta}$ (mas/yr)}                 &
\colhead{$RV$}                  
}
\startdata
HIP 476	&	6.43	&	5.55	&	3.77	&	0.88	&	1.78	&	8.75	&	0.30	&	43.58	&	-5.96	&	\nodata	\\
HIP 3031	&	5.24	&	4.40	&	2.07	&	0.84	&	2.33	&	19.91	&	0.19	&	-229.04	&	-253.11	&	\nodata	\\
HIP 4197	&	7.56	&	6.71	&	4.51	&	0.85	&	2.20	&	4.78	&	1.16	&	-2.54	&	-46.24	&	\nodata	\\
HIP 4960	&	9.31	&	8.60	&	6.58	&	0.71	&	2.03	&	3.18	&	0.93	&	13.62	&	-20.59	&	\nodata	\\
HIP 5104	&	8.40	&	7.65	&	5.59	&	0.75	&	2.06	&	3.19	&	0.79	&	63.61	&	49.81	&	\nodata	\\
HIP 8404	&	6.88	&	5.91	&	3.86	&	0.97	&	2.05	&	10.16	&	0.47	&	-7.40	&	19.26	&	\nodata	\\
HIP 11924	&	8.63	&	8.00	&	6.50	&	0.63	&	1.50	&	11.02	&	0.77	&	-62.76	&	-57.96	&	\nodata	\\
HIP 13339	&	6.75	&	5.87	&	3.85	&	0.88	&	2.02	&	8.34	&	0.37	&	-25.49	&	-24.74	&	\nodata	\\
HIP 19611	&	7.60	&	6.60	&	4.44	&	1.00	&	2.17	&	6.87	&	0.77	&	-11.57	&	14.79	&	4.0	\\
HIP 19740	&	5.71	&	4.89	&	2.97	&	0.82	&	1.92	&	9.83	&	0.64	&	-10.31	&	-30.01	&	\nodata	\\
HIP 27280	&	6.67	&	5.79	&	3.93	&	0.88	&	1.86	&	10.95	&	0.50	&	-32.08	&	-65.50	&	\nodata	\\
HIP 38801	&	9.14	&	8.48	&	6.83	&	0.66	&	1.65	&	1.64	&	1.35	&	-2.02	&	1.08	&	11.1	\\
HIP 39326	&	6.91	&	6.30	&	4.82	&	0.61	&	1.48	&	8.71	&	0.40	&	6.87	&	-10.52	&	\nodata	\\
HIP 44154	&	6.13	&	5.23	&	3.15	&	0.90	&	2.08	&	11.03	&	0.28	&	-43.78	&	-35.03	&	\nodata	\\
HIP 45033	&	6.90	&	6.03	&	4.09	&	0.87	&	1.94	&	8.86	&	0.42	&	2.64	&	5.37	&	\nodata	\\
HIP 45158	&	6.68	&	5.73	&	3.54	&	0.95	&	2.19	&	10.73	&	0.37	&	-56.68	&	45.79	&	\nodata	\\
HIP 45412	&	6.80	&	5.99	&	3.86	&	0.81	&	2.13	&	7.69	&	0.54	&	-158.20	&	47.73	&	\nodata	\\
HIP 46325	&	9.13	&	8.38	&	6.76	&	0.74	&	1.63	&	11.28	&	1.00	&	13.88	&	-150.81	&	\nodata	\\
HIP 51179	&	7.49	&	6.65	&	4.53	&	0.84	&	2.12	&	8.05	&	0.56	&	-38.90	&	14.13	&	-25.2	\\
HIP 51487	&	10.33	&	9.53	&	7.64	&	0.80	&	1.89	&	16.89	&	1.23	&	-6.09	&	-33.43	&	-14.5	\\
HIP 54048	&	7.21	&	6.35	&	4.32	&	0.86	&	2.03	&	6.04	&	0.46	&	5.18	&	-5.06	&	-6.3	\\
HIP 56194	&	7.47	&	6.55	&	4.41	&	0.93	&	2.14	&	5.66	&	0.41	&	-25.08	&	-25.49	&	\nodata	\\
HIP 57535	&	9.28	&	8.60	&	7.06	&	0.68	&	1.54	&	5.51	&	0.86	&	39.71	&	-7.07	&	\nodata	\\
HIP 57748	&	8.66	&	7.91	&	6.01	&	0.75	&	1.90	&	4.47	&	0.74	&	23.31	&	-19.18	&	-22.0	\\
HIP 58269	&	9.41	&	8.70	&	7.18	&	0.71	&	1.51	&	10.32	&	1.12	&	80.71	&	-144.32	&	\nodata	\\
HIP 60140	&	8.50	&	7.67	&	5.60	&	0.83	&	2.07	&	3.75	&	0.79	&	-15.08	&	3.63	&	0.9	\\
HIP 60485	&	5.64	&	4.77	&	2.82	&	0.87	&	1.96	&	8.44	&	0.24	&	12.64	&	11.43	&	\nodata	\\
HIP 60873	&	9.61	&	8.81	&	6.74	&	0.80	&	2.07	&	5.39	&	1.09	&	-58.88	&	2.83	&	-32.7	\\
HIP 62325	&	6.66	&	5.70	&	3.44	&	0.96	&	2.26	&	22.15	&	0.35	&	279.34	&	-453.41	&	\nodata	\\
HIP 65900	&	8.93	&	8.34	&	7.00	&	0.59	&	1.34	&	16.74	&	0.91	&	-132.28	&	-29.61	&	\nodata	\\
HIP 66892	&	7.16	&	6.30	&	4.26	&	0.86	&	2.04	&	3.58	&	0.40	&	-85.88	&	13.24	&	\nodata	\\
HIP 70341	&	9.45	&	8.82	&	7.20	&	0.63	&	1.62	&	14.36	&	1.02	&	-147.37	&	-24.82	&	\nodata	\\
HIP 70344	&	7.80	&	7.22	&	5.79	&	0.58	&	1.43	&	14.20	&	0.78	&	20.49	&	-151.26	&	\nodata	\\
HIP 71837	&	6.50	&	5.56	&	3.41	&	0.94	&	2.15	&	8.09	&	0.33	&	-158.78	&	-112.62	&	\nodata	\\
HIP 72631	&	5.91	&	4.90	&	2.80	&	1.01	&	2.10	&	14.92	&	0.40	&	89.97	&	-124.57	&	\nodata	\\
HIP 75823	&	9.33	&	8.54	&	6.44	&	0.79	&	2.10	&	2.49	&	1.16	&	-23.31	&	9.96	&	\nodata	\\
HIP 78990	&	5.14	&	4.33	&	2.47	&	0.81	&	1.86	&	11.22	&	0.32	&	44.81	&	-45.42	&	\nodata	\\
HIP 80309	&	6.63	&	5.67	&	3.68	&	0.96	&	1.99	&	6.44	&	0.32	&	-29.31	&	34.26	&	\nodata	\\
HIP 80543	&	7.30	&	6.69	&	5.22	&	0.61	&	1.47	&	12.56	&	0.55	&	-0.46	&	-3.42	&	\nodata	\\
HIP 82014	&	9.67	&	8.92	&	7.14	&	0.75	&	1.78	&	8.23	&	0.80	&	-48.14	&	119.81	&	-19.9	\\
HIP 85715	&	6.58	&	5.64	&	3.57	&	0.95	&	2.07	&	8.27	&	0.27	&	3.77	&	16.33	&	\nodata	\\
HIP 89008	&	6.49	&	5.57	&	3.65	&	0.92	&	1.92	&	8.28	&	0.30	&	12.72	&	11.12	&	\nodata	\\
HIP 89095	&	7.98	&	7.13	&	5.12	&	0.85	&	2.01	&	2.73	&	0.51	&	8.25	&	14.35	&	\nodata	\\
HIP 90906	&	8.33	&	7.44	&	5.29	&	0.89	&	2.15	&	3.39	&	0.73	&	4.63	&	14.69	&	\nodata	\\
HIP 91985	&	7.07	&	6.26	&	4.38	&	0.81	&	1.88	&	3.35	&	0.30	&	9.93	&	-2.23	&	\nodata	\\
HIP 92827	&	8.46	&	7.63	&	5.45	&	0.83	&	2.18	&	5.06	&	0.60	&	20.70	&	21.85	&	\nodata	\\
HIP 93940	&	8.79	&	8.25	&	6.88	&	0.54	&	1.37	&	7.81	&	0.83	&	37.43	&	7.16	&	-53.4	\\
HIP 94598	&	6.91	&	6.04	&	4.05	&	0.87	&	1.99	&	4.79	&	0.42	&	3.33	&	-10.99	&	\nodata	\\
HIP 94779	&	4.73	&	3.80	&	1.76	&	0.93	&	2.04	&	26.27	&	0.10	&	60.07	&	122.83	&	\nodata	\\
HIP 98587	&	8.55	&	7.70	&	5.65	&	0.85	&	2.05	&	2.68	&	0.67	&	3.52	&	-26.95	&	\nodata	\\
HIP 100274	&	7.07	&	6.17	&	4.25	&	0.90	&	1.93	&	5.88	&	0.52	&	-3.10	&	6.71	&	\nodata	\\
HIP 103004	&	5.39	&	4.58	&	2.72	&	0.81	&	1.86	&	17.30	&	0.66	&	-75.44	&	-62.03	&	\nodata	\\
HIP 103734	&	6.99	&	6.01	&	4.04	&	0.98	&	1.97	&	5.21	&	0.33	&	-3.47	&	-6.42	&	\nodata	\\
HIP 113610	&	7.10	&	6.23	&	4.22	&	0.87	&	2.00	&	5.57	&	0.69	&	44.21	&	9.99	&	-14.7	\\
HIP 114809	&	7.64	&	6.80	&	4.76	&	0.84	&	2.04	&	6.22	&	0.45	&	-0.29	&	12.71	&	\nodata	\\
HIP 115839	&	7.26	&	6.38	&	4.33	&	0.88	&	2.05	&	7.97	&	0.42	&	99.16	&	-21.66	&	\nodata	\\
HIP 118209	&	5.81	&	4.89	&	2.95	&	0.92	&	1.94	&	13.91	&	0.28	&	-57.13	&	-72.08	&	\nodata	\\
HD 9097	&	10.76	&	10.20	&	8.51	&	0.56	&	1.69	&	\nodata	&	\nodata	&	\nodata	&	\nodata	&	\nodata	\\
HD 84686	&	9.40	&	8.55	&	6.44	&	0.85	&	2.11	&	\nodata	&	\nodata	&	\nodata	&	\nodata	&	\nodata	\\
HD 96780	&	9.31	&	8.80	&	7.52	&	0.51	&	1.28	&	\nodata	&	\nodata	&	\nodata	&	\nodata	&	\nodata	\\
HD 101014	&	10.00	&	9.40	&	8.02	&	0.60	&	1.38	&	\nodata	&	\nodata	&	\nodata	&	\nodata	&	\nodata	\\
HD 141770	&	9.53	&	8.88	&	7.09	&	0.65	&	1.80	&	\nodata	&	\nodata	&	\nodata	&	\nodata	&	\nodata	\\
HD 166310	&	10.25	&	9.54	&	7.66	&	0.71	&	1.88	&	\nodata	&	\nodata	&	\nodata	&	\nodata	&	\nodata	\\
HD 221744	&	9.84	&	9.22	&	7.50	&	0.62	&	1.72	&	\nodata	&	\nodata	&	\nodata	&	\nodata	&	\nodata	\\
HD 235802	&	9.78	&	9.06	&	7.28	&	0.72	&	1.78	&	\nodata	&	\nodata	&	\nodata	&	\nodata	&	\nodata	\\
HD 242647	&	10.31	&	9.70	&	8.23	&	0.61	&	1.47	&	\nodata	&	\nodata	&	\nodata	&	\nodata	&	\nodata	\\
HD 243170	&	10.06	&	9.50	&	8.04	&	0.56	&	1.47	&	\nodata	&	\nodata	&	\nodata	&	\nodata	&	\nodata	\\
BD-14 1413	&	9.84	&	9.30	&	7.87	&	0.54	&	1.43	&	\nodata	&	\nodata	&	\nodata	&	\nodata	&	\nodata	\\
BD+27 2057	&	10.43	&	9.50	&	7.22	&	0.93	&	2.28	&	\nodata	&	\nodata	&	\nodata	&	\nodata	&	\nodata	\\
BD+31 2565	&	10.88	&	10.35	&	9.02	&	0.53	&	1.33	&	\nodata	&	\nodata	&	\nodata	&	\nodata	&	\nodata	\\
BD+32 2190	&	10.64	&	9.99	&	8.55	&	0.65	&	1.45	&	\nodata	&	\nodata	&	\nodata	&	\nodata	&	\nodata	\\
BD+41 2221	&	10.45	&	9.82	&	8.17	&	0.63	&	1.65	&	\nodata	&	\nodata	&	\nodata	&	\nodata	&	\nodata	\\
BD+45 1958	&	10.74	&	10.01	&	8.33	&	0.73	&	1.68	&	\nodata	&	\nodata	&	\nodata	&	\nodata	&	\nodata	\\
BD+45 2032	&	10.56	&	10.00	&	8.22	&	0.56	&	1.78	&	\nodata	&	\nodata	&	\nodata	&	\nodata	&	\nodata	\\
BD+54 2710	&	11.10	&	9.90	&	7.05	&	1.20	&	2.85	&	\nodata	&	\nodata	&	\nodata	&	\nodata	&	\nodata	\\
TYC 3720-324-1	&	11.30	&	10.50	&	8.54	&	0.80	&	1.96	&	\nodata	&	\nodata	&	\nodata	&	\nodata	&	\nodata
\enddata

\end{deluxetable}

\end{center}

\clearpage

\begin{center}
\begin{deluxetable}{lcccc}
\tabletypesize{\footnotesize}
\tablewidth{0pt}
\tablecaption{S/N ratios.\label{tab-ston}}
\tablecolumns{5}
\tablehead{
\colhead{Stars}     & \colhead{V} & \multicolumn{3}{c}{S/N} \\
\cline{3-5}
\colhead{} & \colhead{} & \colhead{4500 {\AA}}   & \colhead{5500 {\AA}}   & \colhead{6500 {\AA}}
}
\startdata
BD+27 2057 & 10.43 & 120 & 170 & 160 \\
HD 84686 & 9.4 & 95 & 200 & 190 \\
HIP 98587 & 8.55 & 115 & 250 & 210
\enddata

\end{deluxetable}

\end{center}

\clearpage

\begin{center}
\begin{deluxetable}{lcccr}
\tabletypesize{\footnotesize}
\tablewidth{0pt}
\tablecaption{Model Atmosphere Parameters\label{tab-model}}
\tablecolumns{5}
\tablehead{
\colhead{Star}           &
\colhead{T$_{eff}$}              &
\colhead{log $g$}           &
\colhead{$v_{t}$}     &
\colhead{[Fe/H]}     \\
\colhead{}     &
\colhead{(K)}     &
\colhead{}     &
\colhead{(km s$^{-1}$)}
}
\startdata
HIP 476	&	5140	&	2.80	&	1.35	&	-0.03	\\
HIP 3031	&	5020	&	2.60	&	1.45	&	-0.53	\\
HIP 4197	&	4800	&	2.50	&	1.00	&	-0.23	\\
HIP 4960	&	5260	&	2.55	&	1.60	&	-1.02	\\
HIP 5104	&	5100	&	2.50	&	1.55	&	-0.86	\\
HIP 8404	&	4800	&	2.35	&	1.20	&	-0.05	\\
HIP 11924	&	5850	&	4.00	&	1.00	&	0.30	\\
HIP 13339	&	5120	&	2.85	&	0.95	&	0.23	\\
HIP 19611	&	5080	&	2.90	&	1.25	&	0.00	\\
HIP 19740	&	5100	&	2.40	&	1.20	&	-0.03	\\
HIP 27280	&	5100	&	2.95	&	1.10	&	0.05	\\
HIP 38801	&	5740	&	2.70	&	2.10	&	-0.06	\\
HIP 39326	&	5750	&	3.15	&	1.60	&	-0.21	\\
HIP 44154	&	5000	&	2.60	&	1.20	&	0.04	\\
HIP 45033	&	5120	&	2.75	&	1.10	&	0.02	\\
HIP 45158	&	4900	&	2.70	&	1.40	&	-0.23	\\
HIP 45412	&	5080	&	2.55	&	1.55	&	-0.68	\\
HIP 46325	&	5650	&	4.15	&	1.00	&	0.46	\\
HIP 51179	&	5020	&	2.80	&	1.10	&	-0.15	\\
HIP 51487	&	5440	&	4.55	&	1.30	&	-0.17	\\
HIP 54048	&	5100	&	2.65	&	1.20	&	0.00	\\
HIP 56194	&	4970	&	2.70	&	1.20	&	0.15	\\
HIP 57535	&	5700	&	3.70	&	1.30	&	0.10	\\
HIP 57748	&	5320	&	2.50	&	1.75	&	-0.08	\\
HIP 58269	&	5820	&	4.30	&	1.10	&	0.35	\\
HIP 60140	&	4930	&	2.65	&	0.85	&	-0.08	\\
HIP 60485	&	5210	&	2.65	&	1.65	&	0.08	\\
HIP 60873	&	5080	&	3.25	&	0.90	&	-0.47	\\
HIP 62325	&	4710	&	2.85	&	0.90	&	-0.06	\\
HIP 65900	&	6090	&	4.40	&	1.00	&	0.15	\\
HIP 66892	&	5170	&	2.70	&	1.35	&	-0.02	\\
HIP 70341	&	5700	&	4.40	&	0.90	&	0.00	\\
HIP 70344	&	5950	&	3.70	&	1.25	&	-0.04	\\
HIP 71837	&	4900	&	2.60	&	1.10	&	-0.15	\\
HIP 72631	&	4800	&	2.60	&	1.20	&	-0.21	\\
HIP 75823	&	5200	&	2.85	&	1.10	&	-0.29	\\
HIP 78990	&	5380	&	2.65	&	1.55	&	0.10	\\
HIP 80309	&	5000	&	2.75	&	1.50	&	-0.06	\\
HIP 80543	&	5920	&	3.65	&	1.40	&	0.15	\\
HIP 82014	&	5490	&	3.90	&	1.00	&	0.13	\\
HIP 85715	&	4950	&	2.50	&	1.50	&	0.00	\\
HIP 89008	&	5000	&	2.55	&	1.15	&	0.14	\\
HIP 89095	&	5270	&	2.40	&	1.55	&	0.14	\\
HIP 90906	&	4930	&	2.35	&	1.20	&	0.00	\\
HIP 91985	&	5290	&	2.30	&	1.60	&	-0.06	\\
HIP 92827	&	4850	&	2.70	&	0.85	&	0.04	\\
HIP 93940	&	6110	&	3.80	&	1.30	&	0.20	\\
HIP 94598	&	5010	&	2.45	&	1.30	&	-0.05	\\
HIP 94779	&	4880	&	2.60	&	1.15	&	0.19	\\
HIP 98587	&	4930	&	2.60	&	1.00	&	-0.16	\\
HIP 100274	&	5010	&	2.65	&	1.15	&	0.08	\\
HIP 103004	&	5250	&	2.80	&	1.35	&	-0.09	\\
HIP 103734	&	5060	&	2.65	&	1.25	&	0.31	\\
HIP 113610	&	5140	&	2.55	&	1.45	&	0.00	\\
HIP 114809	&	5120	&	2.65	&	1.20	&	-0.26	\\
HIP 115839	&	5000	&	2.80	&	1.30	&	-0.40	\\
HIP 118209	&	5090	&	2.90	&	1.25	&	0.15	\\
HD 9097	         &	5540	&	4.10	&	0.80	&	0.18	\\
HD 96780  	&	6200	&	4.00	&	1.30	&	0.01	\\
HD 84686 	&	5180	&	2.60	&	1.75	&	-0.05	\\
HD 101014	&	5880	&	4.20	&	0.80	&	0.30	\\
HD 141770	&	5650	&	4.05	&	1.05	&	0.09	\\
HD 166310	&	5480	&	4.60	&	1.35	&	0.17	\\
HD 221744	&	5850	&	4.40	&	1.05	&	0.00	\\
HD 235802	&	5680	&	4.35	&	1.25	&	0.25	\\
HD 242647	&	5890	&	4.30	&	0.95	&	-0.20	\\
HD 243170	&	6100	&	3.70	&	1.10	&	0.15	\\
BD-14 1413	&	5900	&	3.20	&	1.60	&	-0.10	\\
BD+27 2057	&	4810	&	2.25	&	1.25	&	-0.51	\\
BD+31 2565	&	6150	&	4.30	&	1.15	&	0.03	\\
BD+32 2190	&	5860	&	4.25	&	0.80	&	0.29	\\
BD+41 2221	&	5810	&	4.15	&	1.15	&	0.22	\\
BD+45 1958	&	5570	&	4.40	&	0.90	&	0.24	\\
BD+45 2032	&	5420	&	3.70	&	0.80	&	-0.16	\\
BD+54 2710	&	4840	&	2.20	&	1.45	&	0.04	\\
TYC 3720-324-1	&	5960	&	4.20	&	1.05	&	0.38
\enddata

\end{deluxetable}

\end{center}

\clearpage

\begin{center}
\begin{deluxetable}{lcc}
\tabletypesize{\footnotesize}
\tablewidth{0pt}
\tablecaption{The adopted solar abundances.\label{tab-sun}}
\tablecolumns{3}
\tablehead{
\colhead{Species}     &
\colhead{log$\epsilon$(X)}           &
\colhead{$\sigma$}   \\
\colhead{} &
\colhead{(dex)}  &
\colhead{(dex)}
}
\startdata
\ion{Li}{1} & 1.02 & \\
\ion{C}{1} &  8.53 & 0.06  \\
\ion{C}{1} (CH 4300) &  8.38 &  \\
\ion{N}{1} (CN 8000-8040) & 8.2 & \\
\ion{O}{1} (7774) & 8.91 & 0.01 \\
\ion{O}{1} (6300) & 8.64 & \\
\ion{Si}{1} & 7.55 & 0.14  \\
\ion{Ca}{1} & 6.24 & 0.08 \\
\ion{Fe}{1} & 7.5 & 0.08 \\
\ion{Fe}{2} & 7.5 & 0.07 \\
\ion{La}{2} & 1.17 & \\
\ion{Eu}{2} & 0.52 &
\enddata

\end{deluxetable}

\end{center}

\clearpage
\begin{center}
\begin{deluxetable}{lcc}
\tabletypesize{\footnotesize}
\tablewidth{0pt}
\tablecaption{\ciso~ratios and their uncertainties \label{tab-CN}}
\tablecolumns{3}
\tablehead{
\colhead{Stars}              &
\colhead{\ciso}                  &
\colhead{$\sigma$}
}
\startdata
HIP 45412	&	5	&	 +2/-1.5	\\
HIP 115839	&	5	&	 +2/-1	\\
BD+27 2057	&	5	&	 +2/-1	\\
HIP 57748	&	5	&	 +2/-1	\\
HIP 4960           &       6$^a$       &        \nodata \\
HIP 5104	         &	8	&	 +2/-2	\\
HIP 45158	&	9	&	 +3/-1	\\
HIP 45033	&	10	&	 +2/-2	\\
HIP 54048	&	10	&	 +2/-1.5	\\
HD 84686	         &	10	&	 +4/-2	\\
HIP 72631	&	12	&	 +3/-2	\\
HIP 13339	&	12	&	 +5/-2	\\
HIP 19740	&	14	&	 +4/-2	\\
HIP 3031	         &	15	&	 +5/-5	\\
HIP 71837	&	15	&	 +3/-3	\\
HIP 8404	         &	15	&	 +4/-2	\\
HIP 78990	&	15	&	 +3/-2	\\
HIP 94779	&	15	&	 +3/-2	\\
HIP 44154	&	16	&	 +3/-4	\\
HIP 94598	&	17	&	 +5/-3	\\
HIP 98587	&	18	&	 +5/-2	\\
HIP 100274	&	20	&	 +3/-4	\\
HIP 75823	&	20	&	 +8/-5	\\
HIP 60140	&	20	&	 +8/-4	\\
HIP 103004	&	20	&	 +8/-5	\\
HIP 56194	&	20	&	 +7/-5	\\
HIP 90906	&	20	&	 +5/-5	\\
BD+54 2710	&	20	&	 +5/-3	\\
HIP 113610	&	20	&	 +7/-4	\\
HIP 80309	&	20	&	 +5/-3	\\
HIP 92827	&	22	&	 +8/-4	\\
HIP 4197	         &	22	&	 +4/-3	\\
HIP 118209	&	22	&	 +7/-3	\\
HIP 60485	&	25	&	 +5/-5	\\
HIP 89008	&	25	&	 +7/-5	\\
HIP 85715	&	25	&	 +5/-5	\\
HIP 66892	&	30	&	$>$+10	\\
HIP 62325	&	30	&	$>$+10	\\
HIP 19611	&	30	&	$>$+10	\\
HIP 91985	&	30	&	$>$+10	\\
HIP 51179	&	30	&	$>$+10	\\
HIP 103734	&	30	&	$>$+10	\\
HIP 27280	&	30	&	$>$+10
\enddata
\tablecomments{$^a$ Taken from \cite{grat00}.}
\end{deluxetable}

\end{center}

\clearpage
\begin{center}
\begin{deluxetable}{lrrrrr}
\tabletypesize{\footnotesize}
\tablewidth{0pt}
\tablecaption{Li abundances\label{tab-Li}}
\tablecolumns{6}
\tablehead{
\colhead{Stars}           &
\colhead{\teff}              &
\colhead{\logg}           &
\colhead{[Fe/H]}     &
\colhead{log$\epsilon$(Li)}        &
\colhead{\ciso} 
}
\startdata
HIP 45033	&	5120	&	2.75	&	0.02	&	0.93	&	10	\\
HIP 94598	&	5010	&	2.45	&	-0.05	&	0.96	&	17	\\
BD+54 2710	&	4840	&	2.20	&	0.04	&	0.56	&	20	\\
HIP 92827	&	4850	&	2.70	&	0.04	&	0.74	&	22	\\
HIP 4197	&	4800	&	2.50	&	-0.23	&	0.60	&	22	\\
HIP 89008	&	5000	&	2.55	&	0.14	&	1.04	&	25	\\
HIP 103734	&	5060	&	2.65	&	0.31	&	1.34	&	$>$30	\\
HIP 19611	&	5080	&	2.90	&	0.00	&	1.56	&	$>$30	\\
HIP 51179	&	5020	&	2.80	&	-0.15	&	0.75	&	$>$30	\\
HIP 476	&	5140	&	2.80	&	-0.03	&	0.91	&	\nodata	\\
HIP 60873	&	5080	&	3.25	&	-0.47	&	0.81	&	\nodata	\\
BD+45 2032	&	5420	&	3.70	&	-0.16	&	1.54	&	\nodata	\\
HIP 82014	&	5490	&	3.90	&	0.13	&	2.04	&	\nodata	\\
HIP 80543	&	5920&	3.65	&	0.15	&	3.26	&	\nodata	\\
HIP 11924	&	5850	&	4.00	&	0.30	&	1.70	&	\nodata	\\
HD 96780	&	6200	&	4.00	&	0.01	&	2.81	&	\nodata	\\
HD 141770	&	5650	&	4.05	&	0.09	&	2.26	&	\nodata	\\
BD+41 2221	&	5810	&	4.15	&	0.22	&	2.44	&	\nodata	\\
HIP 46325	&	5650	&	4.15	&	0.46	&	1.92	&	\nodata	\\
TYC 3720-324-1	&	5960	&	4.20	&	0.38	&	2.69	&	\nodata	\\
HD 101014	&	5880	&	4.20	&	0.30	&	2.20	&	\nodata	\\
BD+32 2190	&	5860	&	4.25	&	0.29	&	1.59	&	\nodata	\\
HIP 58269	&	5820	&	4.30	&	0.35	&	2.06	&	\nodata	\\
BD+31 2565	&	6150	&	4.30	&	0.03	&	2.57	&	\nodata	\\
HD 242647	&	5890	&	4.30	&	-0.20	&	2.09	&	\nodata	\\
HIP 65900	&	6090	&	4.40	&	0.15	&	2.72	&	\nodata	\\
HD 166310	&	5480	&	4.60	&	0.17	&	2.03	&	\nodata	\\
\enddata

\end{deluxetable}

\end{center}

\clearpage

\begin{center}
\begin{deluxetable}{lrrrrrrrrrc}
\tabletypesize{\footnotesize}
\rotate
\tablewidth{0pt}
\tablecaption{Abundances\label{tab-abun}}
\tablecolumns{11}
\tablehead{
\colhead{Stars}           &
\colhead{[\ion{Si}{1}/Fe]}              &
\colhead{[\ion{Ca}{1}/Fe]}           &
\colhead{[\ion{C}{1}/Fe]$_{CH}$}     &
\colhead{[\ion{N}{1}/Fe]}     &
\colhead{[\ion{O}{1}/Fe]}     &
\colhead{[\ion{O}{1}/Fe]$_{NLTE}$}     &
\colhead{[La/Fe]}     &
\colhead{[Eu/Fe]}     &
\colhead{\ciso}                  &
\colhead{Evol.Phase}
}
\startdata
\multicolumn{11}{c}{Thick Disk} \\
HIP 3031	&	0.21	&	0.18	&	-0.16	&	0.28	&	0.47	&	0.34	&	0.07	&	0.38	&	15 & RHB	\\
HIP 4960	&	0.27	&	0.24	&	-0.07	&	0.72	&	0.66	&	0.59	&	0.17	&	0.54	&	6 & RHB 	\\
HIP 5104	&	0.31	&	0.31	&	-0.23	&	0.51	&	0.62	&	0.76	&	0.07	&	0.51	&	8 & RHB	\\
HIP 45412	&	0.22	&	0.22	&	-0.16	&	0.25	&	0.43	&	0.36	&	0.02	&	0.36	&	5 & RHB	\\
HIP 58269	&	0.12	&	-0.04	&	-0.01	&	0.39	&	0.07	&	0.00	&	-0.25	&	0.11	&	\nodata & 	MS/SG \\
HIP 62325	&	0.17	&	0.10	&	-0.46	&	0.32	&	-0.04	&	0.10	&	-0.08	&	0.04	&	30 & RGB 	\\
HIP 66892	&	0.10	&	0.08	&	-0.41	&	0.36	&	0.06	&	-0.13	&	0.25	&	0.12	&	30 & ---	\\
HIP 71837	&	0.11	&	0.11	&	-0.42	&	0.44	&	0.15	&	0.21	&	0.02	&	0.21	&	15 & RC/RHB 	\\
HIP 72631	&	0.18	&	0.08	&	-0.24	&	0.26	&	0.30	&	0.25	&	-0.01	&	0.25	&	12 & RC	\\
HIP 115839	&	0.27	&	0.24	&	-0.10	&	0.16	&	0.55	&	0.41	&	0.03	&	0.43	&	5 & RHB	\\
BD+27 2057	&	0.31	&	0.30	&	-0.32	&	0.07	&	0.35	&	0.45	&	-0.07	&	0.22	&	5 & RHB/RC	\\		

\multicolumn{11}{c}{Thin/Thick Disk} \\
HIP 11924	&	0.07	&	0.02	&	-0.06	&	0.18	&	0.12	&	0.10	&	-0.17	&	-0.18	&	\nodata & MS/SG	\\
HIP 46325	&	0.08	&	-0.07	&	-0.09	&	0.15	&	0.04	&	-0.08	&	0.06	&	-0.19	&	\nodata & MS/SG	\\
HIP 57535	&	0.12	&	0.06	&	-0.17	&	0.05	&	0.19	&	0.04	&	-0.10	&	-0.01	&	\nodata & SG	\\
HIP 60873	&	0.09	&	0.14	&	-0.13	&	0.20	&	0.26	&	0.19	&	0.22	&	0.38	&	\nodata & RGB	\\
HIP 70341	&	0.10	&	0.07	&	0.00	&	0.10	& 	\nodata	&	0.21	&	0.04	&	0.24	&	\nodata & MS/SG	\\
HIP 70344	&	0.13	&	0.15	&	-0.16	&	0.15	& 	\nodata	&	0.23	&	0.05	&	0.11	&	\nodata & SG	\\
HIP 75823	&	0.09	&	0.16	&	-0.30	&	0.43	&	0.06	&	0.01	&	0.08	&	0.09	&	20 & RHB	\\
HIP 82014	&	0.03	&	0.02	&	-0.10	&	0.11	&	0.13	&	0.00	&	-0.10	&	0.04	&	\nodata & SG	\\
HIP 98587	&	0.08	&	0.11	&	-0.66	&	0.66	&	-0.16	&	0.04	&	0.10	&	0.08	&	18 & RC/eAGB	\\
																
\multicolumn{11}{c}{Thin Disk} \\
HIP 476	&	0.08	&	0.13	&	-0.25	&	0.33	&	0.04	& 	\nodata	&	0.18	&	0.02	&	\nodata & RGB	\\
HIP 4197	&	0.19	&	0.09	&	-0.68	&	0.48	&	0.01	&	0.17	&	0.04	&	0.06	&	22 & RC/RGB	\\
HIP 8404	&	0.11	&	0.11	&	-0.45	&	0.05	&	-0.16	&	-0.22	&	0.06	&	-0.07	&	15 & RC \\	
HIP 13339	&	-0.04	&	0.08	&	-0.39	&	0.03	&	-0.23	&	-0.30	&	0.13	&	-0.04	&	12 & RHB	\\
HIP 19611	&	0.04	&	0.12	&	-0.31	&	0.03	&	-0.13	&	-0.05	&	0.14	&	0.05	&	30 & RGB/RHB \\	
HIP 19740	&	0.02	&	0.12	&	-0.47	&	0.41	&	-0.17	&	-0.28	&	0.16	&	-0.01	&	14 & RHB	\\
HIP 27280	&	0.02	&	0.10	&	-0.38	&	0.26	&	-0.12	&	-0.14	&	0.15	&	0.00	&	30 & RGB/RHB 	\\
HIP 38801	&	0.04	&	0.13	&	-0.37	&	0.63	&	0.11	&	0.22	&	0.29	&	0.24	&	\nodata	& HB \\
HIP 39326	&	0.10	&	0.21	&	-0.18	&	0.43	&	0.23	&	0.36	&	0.01	&	0.09	&	\nodata & SG	\\
HIP 44154	&	0.08	&	0.08	&	-0.45	&	0.19	&	-0.20	&	-0.20	&	0.14	&	0.04	&	16 & RHB	\\
HIP 45033	&	-0.01	&	0.10	&	-0.44	&	0.33	&	-0.18	&	-0.14	&	0.13	&	-0.01	&	10 & RHB	\\
HIP 45158	&	0.14	&	0.07	&	-0.35	&	0.36	&	0.16	&	0.09	&	0.26	&	0.25	&	9 & RC	\\
HIP 51179	&	0.09	&	0.11	&	-0.47	&	0.46	&	-0.08	&	0.05	&	0.13	&	0.00	&	30 & RHB 	\\
HIP 51487	&	0.06	&	0.15	&	-0.04	&	0.14	&	 \nodata	&	0.23	&	0.29	&	-0.06	&	\nodata & MS	\\
HIP 54048	&	0.04	&	0.11	&	-0.59	&	0.37	&	-0.11	&	-0.19	&	0.15	&	0.05	&	10 & RHB  \\	
HIP 56194	&	0.10	&	0.04	&	-0.48	&	0.35	&	-0.12	&	-0.15	&	0.06	&	0.01	&	20 & RC/RGB	\\
HIP 57748	&	0.10	&	0.02	&	-0.20	&	0.09	&	0.02	&	0.11	&	-0.30	&	-0.11	&	6 & RHB	\\
HIP 60140	&	0.05	&	0.02	&	-0.49	&	0.18	&	-0.25	&	-0.10	&	0.16	&	0.11	&	20 & RHB/RC	\\
HIP 60485	&	0.06	&	0.08	&	-0.47	&	0.52	&	-0.08	&	-0.06	&	0.20	&	0.00	&	25 & ---	\\
HIP 65900	&	0.03	&	0.08	&	-0.03	&	\nodata	&	\nodata	&	0.01	&	0.07	&	0.19	&	\nodata & MS 	\\
HIP 78990	&	0.02	&	0.16	&	-0.42	&	0.50	&	-0.03	&	-0.24	&	0.16	&	-0.06	&	15 & RHB	\\
HIP 80309	&	0.14	&	0.11	&	-0.33	&	0.47	&	-0.03	&	-0.12	&	0.03	&	0.08	&	20 & RGB	\\
HIP 80543	&	0.10	&	0.09	&	0.01	&	0.04	& 	 \nodata	 &	0.09	&	-0.05	&	-0.07	&	\nodata & SG 	\\
HIP 85715	&	0.12	&	0.03	&	-0.42	&	0.41	&	-0.03	&	-0.05	&	0.08	&	-0.06	&	25 & RC/RHB	\\
HIP 89008	&	0.05	&	0.06	&	-0.36	&	0.04	&	-0.31	&	-0.28	&	0.00	&	-0.17	&	25 & 	RGB/RHB\\
HIP 89095	&	0.10	&	0.14	&	-0.40	&	0.40	&	-0.19	&	-0.15	&	-0.11	&	-0.06	&	\nodata & HB?	\\
HIP 90906	&	0.14	&	0.09	&	-0.59	&	0.51	&	-0.21	&	-0.06	&	-0.10	&	-0.12	&	20 & RC/RHB	\\
HIP 91985	&	0.05	&	0.12	&	-0.41	&	0.62	&	-0.19	&	-0.15	&	0.06	&	-0.01	&	30 & HB?	\\
HIP 92827	&	0.13	&	0.00	&	-0.75	&	0.72	&	-0.24	&	-0.10	&	-0.12	&	-0.09	&	22 & RC 	\\
HIP 93940	&	0.12	&	0.09	&	-0.10	&	 \nodata	&	-0.02	&	0.11	&	-0.14	&	0.07	&	\nodata & SG 	\\
HIP 94598	&	0.10	&	0.09	&	-0.63	&	0.49	&	-0.13	&	-0.15	&	0.11	&	0.00	&	17 & 	RHB/eAGB\\
HIP 94779	&	0.09	&	0.04	&	-0.53	&	0.14	&	-0.28	&	-0.15	&	-0.10	&	-0.06	&	15 & RC/RHB 	\\
HIP 100274	&	0.02	&	0.05	&	-0.42	&	0.24	&	-0.05	&	-0.09	&	0.06	&	0.02	&	20 & RC/RGB	\\
HIP 103004	&	0.06	&	0.10	&	-0.26	&	0.50	&	-0.10	&	0.11	&	0.12	&	0.13	&	20 & RHB	\\
HIP 103734	&	0.05	&	-0.01	&	-0.34	&	0.28	&	-0.16	&	-0.23	&	-0.06	&	-0.12	&	30 & RGB 	\\
HIP 113610	&	0.08	&	0.13	&	-0.56	&	0.45	&	-0.08	&	-0.16	&	0.13	&	0.02	&	20 & RHB	\\
HIP 114809	&	0.13	&	0.14	&	-0.27	&	0.32	&	0.11	&	0.16	&	0.05	&	0.11	&	\nodata & RHB	\\
HIP 118209	&	0.03	&	0.06	&	-0.42	&	0.25	&	-0.15	&	-0.29	&	0.17	&	0.08	&	22 & RGB/RHB	\\
HD 9097	&	0.05	&	-0.04	&	-0.06	&	0.11	&	 \nodata	&	0.17	&	0.01	&	0.25	&	\nodata & MS/SG	\\
HD 84686	&	0.13	&	-0.09	&	-0.10	&	0.15	&	0.14	&	0.05	&	-0.26	&	0.24	&	10 & RHB	\\
HD 96780  	&	0.12	&	0.14	&	-0.04	&	\nodata	&	0.18	&	0.15	&	0.04	&	0.08	&	\nodata & MS/SG	\\
HD 101014	&	0.04	&	-0.02	&	-0.13	&	0.20	&	\nodata	&	0.04	&	0.00	&	-0.06	&	\nodata & MS/SG	\\
HD 141770	&	0.03	&	0.00	&	-0.04	&	0.20	&	0.03	&	0.15	&	0.03	&	0.06	&	\nodata & MS/SG	\\
HD 166310	&	0.04	&	0.02	&	-0.39	&	0.14	&	\nodata	&	0.15	&	0.02	&	0.24	&	\nodata & MS	\\
HD 221744	&	-0.01	&	0.13	&	-0.15	&	0.35	&	0.14	&	0.04	&	0.12	&	0.14	&	\nodata & MS	\\
HD 235802	&	0.12	&	0.02	&	-0.26	&	0.22	&	\nodata	&	0.20	&	\nodata	&	-0.17	&	\nodata & MS	\\
HD 242647	&	0.02	&	0.11	&	-0.17	&	0.11	&	\nodata	&	0.33	&	0.17	&	0.20	&	\nodata  & MS	\\
HD 243170	&	0.03	&	0.09	&	-0.20	&	\nodata	&	\nodata	&	0.03	&	\nodata	&	-0.03	&	\nodata & SG	\\
BD-14 1413	&	0.19	&	0.19	&	-0.17	&	\nodata	&	\nodata	&	0.35	&	0.00	&	0.14	&	\nodata & 	SG \\
BD+31 2565	&	0.07	&	0.03	&	-0.08	&	\nodata	&	\nodata	&	0.10	&	0.25	&	0.04	&	\nodata & MS	\\
BD+32 2190	&	0.03	&	0.00	&	-0.01	&	0.23	&	\nodata	&	0.05	&	-0.13	&	0.07	&	\nodata & MS/SG	\\
BD+41 2221	&	0.05	&	0.02	&	0.04	&	0.19	&	0.13	&	0.04	&	-0.05	&	0.06	&	\nodata & MS/SG	\\
BD+45 1958	&	0.02	&	-0.02	&	-0.24	&	0.29	&	\nodata	&	0.04	&	-0.14	&	0.02	&	\nodata & MS	\\
BD+45 2032	&	0.08	&	0.12	&	-0.29	&	0.14	&	-0.02	&	0.20	&	0.14	&	0.38	&	\nodata & SG	\\
BD+54 2710	&	0.16	&	0.11	&	-0.60	&	0.52	&	-0.22	&	-0.05	&	0.10	&	-0.01	&	20 & RC	\\
TYC 3720-324-1	&	0.10	&	-0.05	&	-0.05	&	0.16	&	\nodata	&	0.04	&	0.07	&	-0.29	&  	\nodata & MS/SG	
\enddata
\tablecomments{Due to lack of parallax information we were unable to
investigate the kinematical parameters of BD+27 2057 but by taking into  
previous studies, e.g.  \cite{stet87}, \cite{upg62,upg63} and \cite{tau01}, 
we list it as a thick disk member in the table.}
\end{deluxetable}

\end{center}

\clearpage

\begin{center}
\begin{deluxetable}{lrrrrrrc}
\tabletypesize{\footnotesize}
\tablewidth{0pt}
\tablecaption{Kinematics\label{tab-kin}}
\tablecolumns{8}
\tablehead{
\colhead{Stars}           &
\colhead{$U_{\rm{LSR}}$}              &
\colhead{$\sigma_{U_{\rm{LSR}}}$}              &
\colhead{$V_{\rm{LSR}}$}           &
\colhead{$\sigma_{V_{\rm{LSR}}}$}           &
\colhead{$W_{\rm{LSR}}$}     &
\colhead{$\sigma_{W_{\rm{LSR}}}$}     &
\colhead{Membership} \\
\colhead{} &
\colhead{(\kmsec)} &
\colhead{(\kmsec)} &
\colhead{(\kmsec)} &
\colhead{(\kmsec)} &
\colhead{(\kmsec)} &
\colhead{(\kmsec)} &
\colhead{} 
}
\startdata
HIP 62325	&	117.5	&	1.6	&	-55.0	&	0.7	&	29.6	&	0.5	&Tk	\\
HIP 3031	&	110.1	&	0.7	&	-54.6	&	0.7	&	6.9	&	0.7	&	Tk	\\
HIP 5104	&	-108.0	&	29.2	&	12.1	&	2.6	&	28.3	&	11.3	&	Tk	\\
HIP 71837	&	-21.1	&	1.0	&	-106.9	&	4.5	&	5.8	&	1.1	&	Tk	\\
HIP 66892	&	-88.2	&	11.0	&	-61.7	&	6.5	&	-6.9	&	2.8	&	Tk	\\
HIP 72631	&	99.2	&	1.1	&	-13.2	&	0.4	&	38.7	&	1.1	&	Tk	\\
HIP 45412	&	-100.2	&	5.2	&	27.7	&	2.0	&	-22.9	&	5.0	&	Tk	\\
HIP 4960	&	46.7	&	5.2	&	-83.0	&	10.4	&	45.0	&	9.2	&	Tk	\\
HIP 115839	&	-49.8	&	2.5	&	-59.4	&	2.0	&	58.5	&	2.2	&	Tk	\\
HIP 58269	&	74.1	&	7.2	&	-42.2	&	5.2	&	14.5	&	11.3	&	Tk	\\
HIP 82014	&	-57.6	&	7.0	&	-29.8	&	2.0	&	-4.4	&	0.5	&	Tk/Tn	\\
HIP 11924	&	63.3	&	2.1	&	-13.0	&	1.1	&	-3.2	&	2.3	&	Tk/Tn	\\
HIP 46325	&	13.2	&	2.1	&	-61.1	&	5.4	&	15.5	&	1.5	&	Tk/Tn	\\
HIP 70344	&	49.6	&	2.4	&	-29.0	&	2.9	&	1.0	&	6.1	&	Tk/Tn	\\
HIP 98587	&	51.8	&	8.6	&	8.4	&	4.0	&	-23.1	&	7.8	&	Tk/Tn	\\
HIP 70341	&	-24.7	&	1.9	&	-47.7	&	2.7	&	-20.1	&	1.3	&	Tk/Tn	\\
HIP 57535	&	49.4	&	5.1	&	11.9	&	1.4	&	-4.6	&	1.5	&	Tk/Tn	\\
HIP 75823	&	-41.6	&	17.0	&	-25.8	&	8.5	&	3.9	&	13.2	&	Tk/Tn	\\
HIP 60873	&	-27.4	&	9.3	&	-26.9	&	4.9	&	-29.3	&	1.2	&	Tk/Tn	\\
HIP 57748	&	46.1	&	4.8	&	-5.9	&	1.4	&	-1.9	&	1.9	&	Tn	\\
HIP 27280	&	43.9	&	1.9	&	-3.0	&	1.0	&	-14.0	&	1.3	&	Tn	\\
HIP 4197	&	31.2	&	4.0	&	-22.8	&	5.1	&	-24.6	&	9.4	&	Tn	\\
HIP 93940	&	-35.2	&	1.2	&	-24.1	&	1.2	&	-14.7	&	2.0	&	Tn 	\\
HIP 114809	&	1.7	&	0.6	&	42.2	&	1.7	&	-6.3	&	1.2	&	Tn	\\
HIP 118209	&	38.9	&	0.7	&	-5.5	&	0.5	&	1.7	&	0.8	&	Tn	\\
HIP 103004	&	35.4	&	1.2	&	-1.4	&	1.9	&	12.7	&	0.5	&	Tn 	\\
HIP 65900	&	-12.6	&	1.3	&	-31.6	&	1.6	&	-9.4	&	0.5	&	Tn	\\
HIP 89095	&	-31.4	&	5.0	&	-2.7	&	5.1	&	-9.3	&	2.2	&	Tn	\\
HIP 113610	&	-29.0	&	4.6	&	-8.3	&	0.9	&	6.7	&	1.6	&	Tn	\\
HIP 92827	&	-27.7	&	3.6	&	-7.2	&	4.6	&	-9.2	&	1.7	&	Tn	\\
HIP 51179	&	-2.3	&	1.6	&	19.5	&	0.4	&	-22.4	&	0.7	&	Tn	\\
HIP 80309	&	-19.7	&	1.6	&	-22.0	&	1.6	&	0.4	&	1.4	&	Tn	\\
HIP 39326	&	25.7	&	0.4	&	11.9	&	0.4	&	4.2	&	0.4	&	Tn 	\\
HIP 13339	&	26.8	&	1.6	&	1.0	&	1.2	&	-9.0	&	1.0	&	Tn	\\
HIP 94779	&	-17.2	&	0.2	&	-19.4	&	0.9	&	-3.7	&	0.3	&	Tn	\\
HIP 90906	&	-25.0	&	4.0	&	-6.4	&	3.1	&	0.9	&	1.1	&	Tn	\\
HIP 78990	&	15.5	&	0.9	&	3.7	&	0.3	&	-19.7	&	0.8	&	Tn	\\
HIP 45158	&	-19.7	&	1.3	&	15.8	&	1.8	&	0.7	&	0.7	&	Tn	\\
HIP 56194	&	-5.2	&	1.3	&	-20.6	&	2.1	&	13.2	&	1.8	&	Tn	\\
HD 96780	&	-20.1	&	\nodata	&	-13.1	&	\nodata	&	1.9	&	\nodata	&	Tn	\\
HIP 80543	&	-13.4	&	3.3	&	-10.7	&	2.1	&	-16.7	&	3.2	&	Tn	\\
HIP 44154	&	-18.5	&	1.5	&	-12.4	&	0.5	&	5.6	&	1.4	&	Tn	\\
HIP 19740	&	22.5	&	1.0	&	-2.3	&	0.8	&	-0.8	&	1.1	&	Tn	\\
HIP 94598	&	21.3	&	1.4	&	6.8	&	1.7	&	-0.1	&	0.9	&	Tn	\\
HIP 100274	&	12.2	&	1.2	&	14.3	&	1.7	&	10.5	&	0.8	&	Tn	\\
HIP 60485	&	17.1	&	0.4	&	10.2	&	0.4	&	-5.6	&	0.8	&	Tn \\
HIP 91985	&	10.6	&	0.7	&	17.4	&	1.9	&	-3.3	&	1.6	&	Tn	\\
HIP 51487	&	18.3	&	0.4	&	-1.8	&	0.7	&	-6.3	&	0.5	&	Tn	\\
HIP 45033	&	15.2	&	1.4	&	10.3	&	0.7	&	4.4	&	1.3	&	Tn	\\
HIP 8404	&	7.3	&	1.0	&	14.7	&	0.8	&	8.8	&	1.7	&	Tn	\\
HIP 19611	&	4.8	&	0.5	&	17.4	&	1.6	&	4.2	&	0.7	&	Tn	\\
HIP 54048	&	15.9	&	0.7	&	8.8	&	0.4	&	2.2	&	0.4	&	Tn	\\
HIP 103734	&	14.5	&	1.0	&	-5.9	&	4.9	&	6.9	&	0.7	&	Tn	\\
HIP 89008	&	-7.3	&	0.9	&	-10.7	&	1.7	&	-8.2	&	0.9	&	Tn	\\
HIP 85715	&	-11.1	&	1.1	&	-8.1	&	1.4	&	-5.8	&	1.0	&	Tn	\\
HIP 476	&	-9.4	&	0.8	&	-6.4	&	1.4	&	-0.4	&	1.5	&	Tn	\\
HIP 60140	&	-8.7	&	4.1	&	0.2	&	1.7	&	4.2	&	0.9	&	Tn	\\
HIP 38801	&	-2.5	&	4.8	&	2.1	&	4.5	&	6.3	&	5.6	&	Tn
\enddata

\end{deluxetable}

\end{center}

\end{document}